\documentclass[twoside,twocolumn,9pt]{article}
\usepackage{extsizes}
\usepackage[super,sort&compress,comma]{natbib} 
\usepackage[version=3]{mhchem}
\usepackage[left=1.5cm, right=1.5cm, top=1.785cm, bottom=2.0cm]{geometry}
\usepackage{balance}
\usepackage{mathptmx}
\usepackage{sectsty}
\usepackage{graphicx} 
\usepackage{lastpage}
\usepackage{float}
\usepackage{fancyhdr}
\usepackage{fnpos}
\usepackage[english]{babel}
\usepackage{array}
\usepackage{droidsans}
\usepackage{charter}
\usepackage[T1]{fontenc}
\usepackage[usenames,dvipsnames]{xcolor}
\usepackage{setspace}
\usepackage[compact]{titlesec}
\usepackage{hyperref}
\usepackage{stfloats}

\definecolor{cream}{RGB}{222,217,201}

\emergencystretch=\maxdimen
\newcommand{\eg}{{\it e.g.}}
\newcommand{\ie}{{\it i.e.}}

\usepackage{comment}
\usepackage{chngcntr}
\usepackage{booktabs}
\usepackage{multirow}
\usepackage{mathtools, nccmath}

\DeclarePairedDelimiter{\nint}\lfloor\rceil

\begin{document}

\title{\textbf{Real-Time Tilt Undersampling Optimization during Electron Tomography of Beam Sensitive Samples using Golden Ratio Scanning and RECAST3D $^\dag$}} 

\author{\large{Timothy M. Craig,\textit{$^{a}$} Ajinkya A Kadu,\textit{$^{a,b}$} Kees Joost Batenburg,\textit{$^{b,c}$} and Sara Bals$^{\ast}$\textit{$^{a}$}} \\[1ex]
\small{\textit{$^{a}$}~Electron Microscopy for Materials Science and NANOlab Center of Excellence, University of Antwerp, Groenenborgerlaan 171, Antwerp 2020, Belgium.} \\
\small{\textit{$^{b}$}~Centrum Wiskunde \& Informatica, Science Park 123, Amsterdam 1098 XG, The Netherlands} \\
\small{\textit{$^{c}$}~Leiden Institute of Advanced Computer Science,  Leiden University, Niels Bohrweg 1, 2333CA Leiden, The Netherlands.}
} 

\date{}

\maketitle

\begin{abstract}
    Electron tomography is a widely used technique for 3D structural analysis of nanomaterials, but it can cause damage to samples due to high electron doses and long exposure times. To minimize such damage, researchers often reduce beam exposure by acquiring fewer projections through tilt undersampling. However, this approach can also introduce reconstruction artifacts due to insufficient sampling. Therefore, it is important to determine the optimal number of projections that minimizes both beam exposure and undersampling artifacts for accurate reconstructions of beam-sensitive samples. Current methods for determining this optimal number of projections involve acquiring and post-processing multiple reconstructions with different numbers of projections, which can be time-consuming and requires multiple samples due to sample damage. To improve this process, we propose a protocol that combines golden ratio scanning and quasi-3D reconstruction to estimate the optimal number of projections in real-time during a single acquisition. This protocol was validated using simulated and realistic nanoparticles, and was successfully applied to reconstruct two beam-sensitive metal-organic framework complexes.
\end{abstract}




\section{Introduction}
\label{sec:Introduction}

Nanomaterials are materials with at least one dimension in the nanoscale, usually ranging from 1 to 100 nanometers. \cite{roduner2006size} They have unique physical, chemical, and spectroscopic properties compared to their bulk counterparts, which can be used for various commercial, industrial, and medicinal purposes.\cite{laux2018nanomaterials, barreto2011nanomaterials} These properties are largely influenced by the three-dimensional (3D) structure and morphology of the nanomaterial. Therefore, it is essential to accurately characterize the nanomaterial's structure to understand its behavior and predict its potential applications.\cite{roduner2006size, calvaresi2020route, choo2021nanoparticle}

High-resolution imaging techniques such as transmission electron microscopy (TEM) and annular dark-field scanning transmission electron microscopy (ADF-STEM) can provide insights into the structure of nanomaterials.\cite{urban2008studying, crewe1968high, shin1989annular} However, these techniques only produce two-dimensional (2D) projections, which may not accurately represent the true 3D structure of the material. To overcome this limitation, techniques such as electron tomography (ET) have been developed to enable the three-dimensional characterization of nanomaterials.\cite{kubel2005recent, midgley2009electron, scott2012electron}  In a typical ET procedure, a series of 2D images are obtained at incremental angles ($1 - 3^{\circ}$) over a range of approximately $ \pm 70 - 80^{\circ}$ .\cite{bartesaghi2008classification, goris2012electron, pryor2017genfire} These images are then aligned and processed using reconstruction algorithms such as filtered back projection (FBP),\cite{dudgeon1984multidimensional} simultaneous iterative reconstruction technique (SIRT),\cite{gordon1970algebraic} or expectation maximization (EM)\cite{lange1984reconstruction} to generate a 3D volume of the nanomaterial. Overall, the use of ET has significantly improved the scientific understanding of nanomaterials and their potential applications.

Exposure to the electron beam during the tomographic acquisition of nanomaterials can cause significant deformation due to various factors, including radiolysis, atomic displacement, heating, charge accumulation, and knock-on effects. \cite{ugurlu2011radiolysis, crespi1996anisotropic, russo2018charge,  banhart2006irradiation, jiang2015electron} This electron beam-induced damage has been observed in a wide range of nanomaterials, including silicates, zeolites, and metal-organic frameworks (MOFs). \cite{liu2020bulk, zak2021guide, jiang2015electron, turner2008direct, treacy1987electron} Prior studies have investigated various approaches to minimize beam damage, with the most reliable method being limiting beam exposure. \cite{egerton2012mechanisms} Low-dose methods, which employ acquisition regimes with high signal-to-noise ratios,\cite{mcmullan2014comparison} such as ptychography \cite{chen2018biomineralized} or integrated differential phase contrast\cite{liu2020bulk} microscopy, allow for the collection of the same signal with significantly less exposure. However, these techniques require specialized detectors and setups that may not be readily available. Tracking and focusing optimization during acquisition have also been used to reduce beam exposure. For example, in 2018, a twofold reduction in beam exposure was achieved by accelerating the high-angle annular dark-field scanning transmission electron microscopy (HAADF-STEM) acquisition to a few minutes through simultaneous scanning, tracking, and focusing. \cite{vanrompay2021fast,vanrompay20183d}

\begin{figure}[t]
	\centering
	\includegraphics[width=\columnwidth]{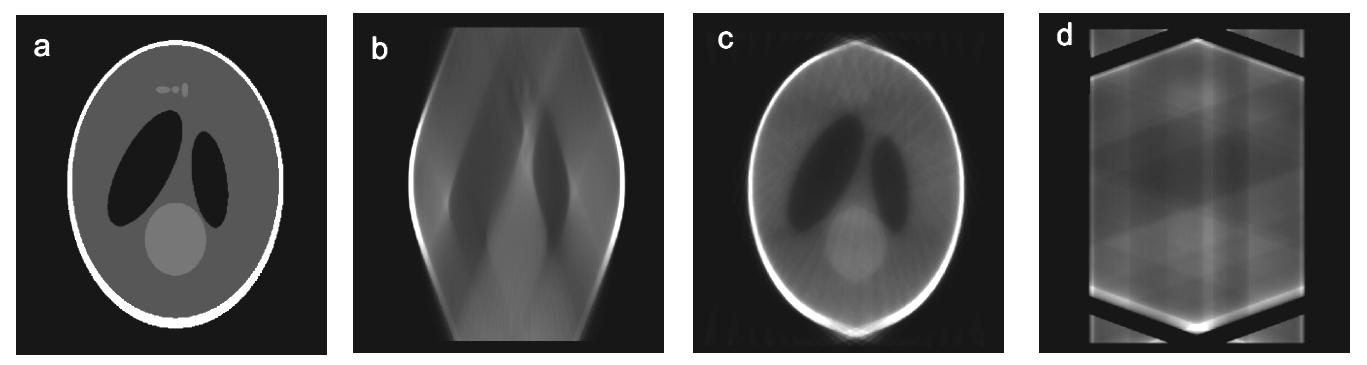}
	\caption{Phantom Shepp-Logan (a) undersampled with 21 projections by reducing the annular range ($ \pm20^\circ$, $2^\circ$ step, 21 projections) (b) and the sampling density ($\pm 70^\circ$, $7^\circ$ step, 21 projections) (c). By decreasing the sampling density further ($\pm 70^\circ$, $70^\circ$ step, three projections), undersampling artefacts become apparent (d).}
    \label{fig:1}
\end{figure}

Another commonly applied technique to reduce beam damage is undersampling. Undersampling reduces beam exposure by reducing (\emph{i}) the information encoded into an image (image undersampling)\cite{cho2012feasibility} or (\emph{ii}) the number of images collected (tilt undersampling). \cite{davison1983ill}. However, undersampling also has its own challenges; typically in the introduction of new artefacts. This has been extensively studied for various undersampling schemes by Vanrompay et. al.\cite{vanrompay2019experimental} For instance, tilt undersampling has been shown to amplify missing wedge artefacts by decreasing the angular range (\eg, from 70° to 15°).\cite{frikel2013characterization, bartesaghi2008classification, vanrompay2019experimental}
Whilst these artefacts can be compensated using algorithms such as discrete algebraic reconstruction tomography (DART), these algorithms assume prior knowledge of the sample's properties, which may not always be applicable. \cite{ batenburg2011dart, goris2012electron, pryor2017genfire}

The missing wedge is an issue that arises in ET when only projections over limited angular range are collected. This can be mitigated during undersampling by evenly distributing the few images across the entire available annular range, thus decreasing the sampling density rather than the sampling range. \cite{vanrompay2019experimental} For instance, both an acquisition at $\pm 20^{\circ}$ in $2^{\circ}$ increments and $\pm 70^{\circ}$ in $7^{\circ}$ increments have 21 projections. However, the missing wedge is minimized in the latter case, where the images are more evenly distributed. Nevertheless, even in this case, significantly decreasing the sampling density can result in artefacts in the reconstructed image (Figure~\ref{fig:1}). Therefore, it is important to find a balance between minimizing beam damage and avoiding undersampling artefacts in order to ensure the quality of the reconstruction.

\begin{figure}[t]
	\centering
	\includegraphics[width=\columnwidth]{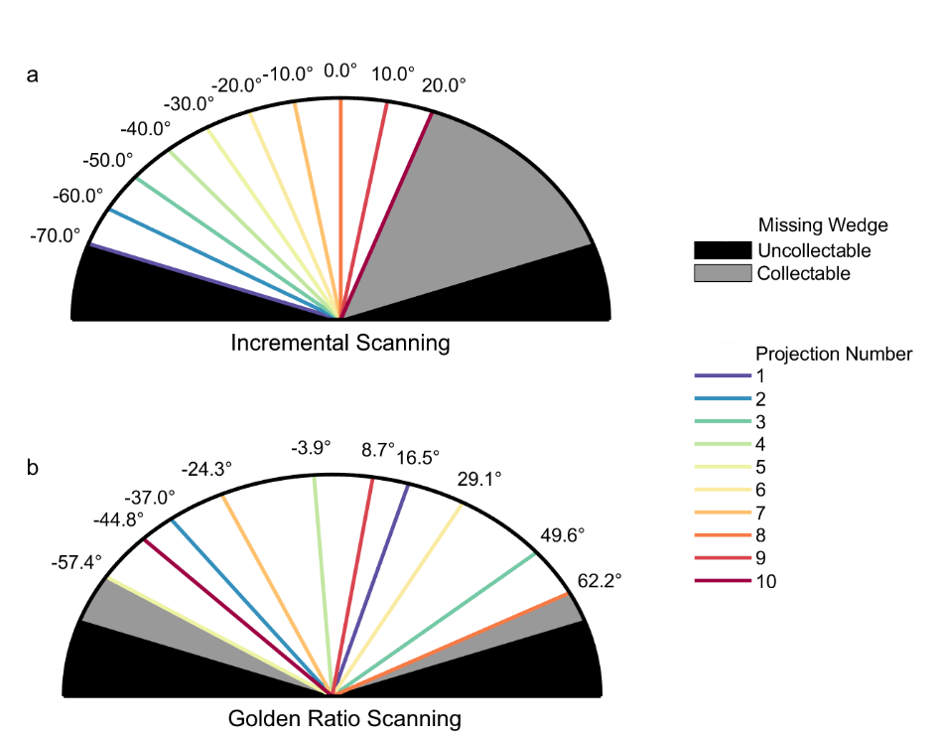}
	\caption{First $10$ projections acquired using IS ($ \pm 70^\circ$, $10^\circ$ increment)(a) and GRS ($\pm 70^\circ$)(b). The collectable missing wedge due to early termination of acquisition (grey) and the inaccessible missing wedge due to the holder geometry (black) are shown for both acquisition schemes.}
    \label{fig:2}
\end{figure}

When assessing the quality of a reconstruction, researchers often compare it to a reference structure collected using a standard $2-3^{\circ}$ tilt increment. \cite{vanrompay2019experimental} However, this approach is not suitable for optimizing tilt undersampling of beam-sensitive samples. Firstly, it is not guaranteed that the reference, collected under standard imaging conditions, does not contain beam damage artefacts. Secondly, during incremental scanning (IS), the sampling density remains constant while the sampling range increases with each new projection (Figure~\ref{fig:2}a). Therefore, prematurely ending the acquisition will result in a large missing wedge in the tilt-series. In order to find the optimal number of projections, the 3D reconstruction of multiple tilt-series collected with different numbers of projections should be compared. This process is time-consuming, as it involves the microscopist alternating between the microscope and post-processing steps at a workstation. \cite{vanrompay2020real} Furthermore, when multiple acquisitions are performed on the same particle, the damage induced in previous acquisitions is often evident in the new tilt-series. Therefore, it is necessary to perform each acquisition on a new particle, making it difficult to directly compare the resulting 3D reconstructions. In such cases, the microscopist must rely on a qualitative judgement to determine the optimal number of projections.

The requirement for multiple tilt-series in tomographic acquisitions can be mitigated using an acquisition scheme with a semi-constant sampling range, and a sampling density that increases as new projections are added. Golden Ratio Scanning (GRS) proposed by Kaestner et al. satisfies this requirement for 4D neutron microtomography.\cite{kaestner2011spatiotemporal} In GRS, the tilt angle $\theta$ in radians is given by
\begin{equation}
\theta = i \, \alpha \,\, \left(\frac{1+\sqrt{5}}{2}\right) \!\!\! \mod(\alpha) - \frac{\alpha}{2},
\label{eq:1}
\end{equation}
where $i$ is the image index, $\mod$ is the modulo function, and $\alpha$ is the annular range in radians. In GRS, the majority of the annular range is occupied within the first $3-4$ projections, and subsequent projections increase the sampling density (Figure~\ref{fig:2}b). Therefore, acquisition can be terminated early without significant missing wedge artefacts. \cite{kaestner2011spatiotemporal} In practice, however, it is impossible to know how many projections are required to minimize undersampling and beam damage artefacts without knowledge of the 3D structure during acquisition.

Quasi-3D reconstruction allows for real-time viewing of 3D data by limiting the computational requirements of reconstruction. This was achieved using the software RECAST3D (Reconstruction of Arbitrary Slices in Tomography), which reduces the computational burden by only reconstructing only a few arbitrary slices at a time using the computationally efficient FBP algorithm. \cite{buurlage2018real}

Here, we present a protocol, referred to as \textit{Tilt Undersampling Optimized Tomographic Acquisition} (TUOTA), that combines GRS with real-time analysis of quasi-3D reconstructions provided by RECAST3D to determine the optimal number of projections for beam-sensitive samples. TUOTA was tested using simulated and experimental datasets, and was applied to two beam-sensitive MOF nanoparticle (NP) composites: NU-1000 encapsulating an Au bipyramidal nanoparticle (Au@NU-1000) and ZIF-8 encapsulating an Au/Pd nanorod (Au/Pd@ZIF-8).

\section{Method}
\label{sec:Method}

\subsection{Tilt Undersampling Optimized Tomographic Acquisition}
\label{sec:Method:TUOTA}
The TUOTA protocol for optimizing the number of projections consists of the following stages:
\begin{enumerate}

\begin{figure*}[b!]
	\centering
	\includegraphics[width=0.8\textwidth]{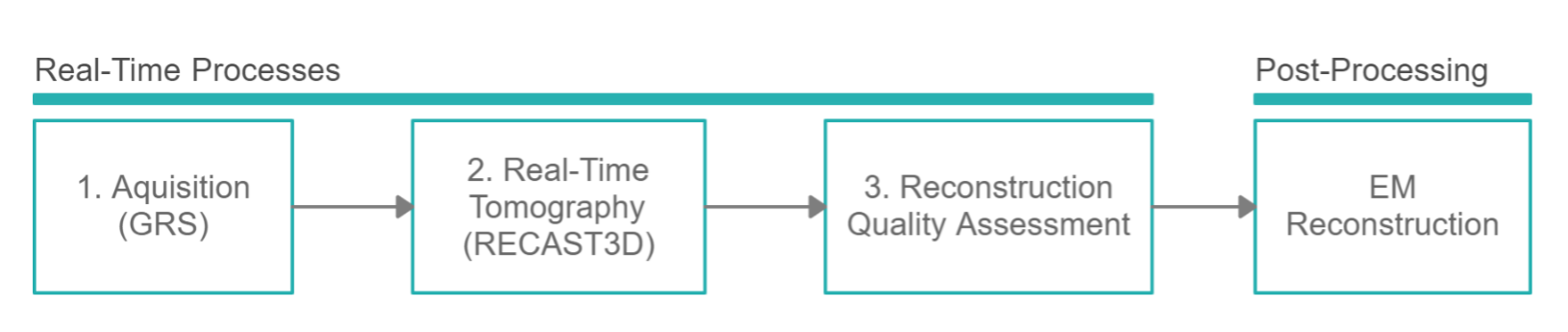}
	\caption{Procedure to optimize the number of projections during 3D reconstruction using TUOTA. Steps 1-3 are performed at the microscope, while the final step can be performed at the compute station (e.g. high-performance computer or server).}
    \label{fig:3}
\end{figure*}

\item Obtaining projections using GRS with an annular range of $\pm 70^\circ$ ($\alpha = 7 \pi / 9$) in real-time at the microscope, with acquisition ending at the discretion of the microscopist.
\item Processing the projections using RECAST3D to reconstruct three slices using FBP, with slices being updated as new projections are acquired.
\item Evaluating slice quality quantitatively based on the number of projections used.
\item Conducting a final reconstruction using the EM algorithm with the optimal number of projections determined in step 3.
\end{enumerate}

\subsubsection{Quantification.~~} 
\label{sec:Method:TUOTA:Quantification}
In order to determine the optimal number of projections, the reconstruction quality of the slices computed in step 2 is quantitatively assessed in step 3. A reliable approach to assess reconstruction quality is to compare it to a reliable reference standard, which is commonly computed as the shape error ($E_s$) or the normalized root-mean-squared difference between the Otsu threshold\cite{otsu1979threshold} binarized reconstruction ($V_{\text{rec}}$) and reference ($V_{\text{ref}}$), defined as
\begin{equation}
	E_s = 100 \,\, \frac{\| V_{\text{ref}} - V_{\text{rec}} \| }{\| V_{\text{ref}} \|} ,
    \label{eq:2}
\end{equation}
where $\| \cdot \|$ represents the Euclidean norm, \ie ,  $\| x \| = \sqrt{ \sum_{i=1}^{n} x_i^2 }$. The reference is typically the sample collected using standard tomographic acquisition, which is assumed to be a reliable representation of the true 3D structure. However, due to beam damage, this assumption may not always be reliable. Additionally, subsequent acquisitions of the same particle may differ from the reference solely due to beam damage induced during the reference acquisition, making it impossible to obtain a reliable reference structure.

In this case, the only information available from RECAST3D is three arbitrary slices. For convenience, in this paper, all calculations were determined from the $xy$, $yz$, and $xz$ orthoslices passing through the origin. If the positions of these orthoslices are fixed for the acquisition duration, the change in these orthoslices can be observed as a function of the number of projections. In ET, as more projections are provided, the reconstruction converges towards a 3D structure, and each projection becomes a smaller portion of the complete set of $N$ projections. Therefore, each projection contributes less to the reconstruction as more projections are added, and the difference between the 3D reconstruction with $N$ and $N-1$ projections tends towards zero.

Applied to the RECAST3D orthoslices, a measure for the convergence can be obtained as a function of $N$ by finding the normalized root mean squared difference between the set of orthoslices ($O_N$) and the orthoslices obtained with $N-1$ projections ($O_{N-1}$). This measure is
\begin{equation}
    \text{SROD}(N) = \frac{\| O_N - O_{N-1} \|}{ \| O_{N} \| }.
    \label{eq:3}
\end{equation}
This metric, referred to as the self-referential orthoslice difference (SROD), can be obtained solely from the RECAST3D orthoslices without a known accurate reference structure. The lower the SROD, the more closely $O_{N-1}$ and $O_N$ resemble each other. Sufficient convergence for reconstruction is achieved when the SROD is lower than a user-defined threshold value. For this work, an arbitrary threshold of $0.1$ was applied. Higher or lower threshold values may be utilized depending on the desired frequency resolution.

The SROD metric only monitors convergence and undersampling. For beam-sensitive samples, beam-induced artifacts may reduce the reconstruction quality before adequately sampling the structure. To monitor this, the signal-to-noise ratio (SNR) of each set of orthoslices is measured as a function of the number of projections, \ie, 
\begin{equation}
	\text{SNR}(N) = 20 \log_{10} \left( \frac{\mu (O_N)}{\sigma (O_N) } \right), 
    \label{eq:4}
\end{equation}
where $\mu$ is the average and $\sigma$ is the standard deviation in the signal of each pixel in the set of orthoslices $O_N$.

It is noted that the SNR typically increases as more projections are added to the tilt-series and tends to decrease in electron microscopy images as a response to beam damage.\cite{heymann2022progressive} The optimum number of projections is determined by the intersection of the convergence of the SROD and the decline in SNR due to beam damage.

The optimum number of projections for a given sample can be obtained by analyzing the SROD and SNR curves as a function of the number of projections. This allows for the determination of the optimum number of projections for a given sample without the need for a reliable reference structure. Refer to the supporting information regarding its implementation in code. 

\subsubsection{Post-processing.~~} The optimal number of projections for the tilt-series is used to reconstruct the complete 3D volume using the MATLAB ASTRA implementation of the EM algorithm. \cite{van2016fast,van2015astra,palenstijn2011performance} In contrast, RECAST3D only provides orthoslices using the FBP algorithm, which has been shown to perform poorly when the tilt-series is undersampled or contains a missing wedge. \cite{van2016fast} Therefore, we prefer the EM algorithm for complete volume reconstruction.

\subsection{Method evaluation}
\label{sec:Method:Evaulation}
To evaluate the validity of TUOTA, we compared the suggested optimum reconstructions to a standardized method for evaluating reconstruction accuracy, $E_s$. However, $E_s$ is not reliable for beam-sensitive samples because the reference sample is unreliable. Therefore, we performed simulated beam damage experiments where a phantom was used as an accurate reference of the initial structure before beam damage was applied. For beam-insensitive samples, it can be assumed that standard IS tomography provides a reasonably accurate reconstruction that can be used as a reference. To evaluate the proposed acquisition procedure and the reliability of TUOTA, we compared the TUOTA- and $E_s$-determined optimum number of projections for both simulated and experimental structures.

\subsubsection{Simulations and Experimental Acquisition.~~} Sample data were obtained through both microscopy simulations and experiments. The simulations were performed by iteratively deforming an original 3D structure using a Gaussian filter and a binomial probability mask implemented in MATLAB. After each iteration, the entire volume was saved. Tilt-series were simulated by forward projecting the structure after each iteration of beam damage. As a result, the image corresponding to the first angle in the tilt-series was simulated by the forward projection of the structure after one iteration of beam damage, and the image for the second angle was obtained by forward projecting the structure after two iterations of beam damage. This process was repeated until images for all angles were obtained. The magnitude of the beam damage was adjustable by adjusting two deformation parameters ($\beta_1, \beta_2$). See the supporting information for more details on the beam damage simulations.

\begin{figure}[t]
	\centering
	\includegraphics[width=\columnwidth]{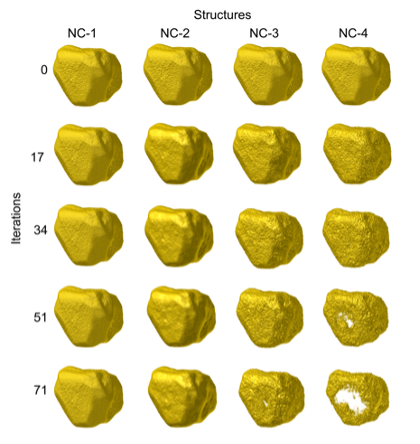}
	\caption{Simulated beam damage on four nanocage samples with different simulation settings. A tilt-series is acquired for each sample by forward projecting after each iteration of simulated deformation. Each nanocage is shown after 0, 17, 34, 51 and 71 iterations of deformation.}
    \label{fig:4}
\end{figure}

Simulations were performed for a nanoparticle with a hollow, cage-like structure, which we previously investigated experimentally. \cite{goris2014monitoring} We used four different deformation settings to simulate the beam damage (Figure~\ref{fig:4}, Movie S1), ranging from no deformation (NC-1) to severe deformation (NC-4). Beam damage in the hollow nanocages manifested as a slowly opening cavity in the structure and thinning of the cage walls.

In addition, three experimental samples were characterized (Figure~B1.): an Au/Pd nanostar (NS) and two NP@MOF composites. The first composite was a \ce{Zn(2-methyl imidazole)2} MOF ZIF-8 containing an Au/Pd nanorod (Au/Pd@ZIF-8), and the second composite was a NU-1000 MOF consisting of \ce{Zr6O4(OH)4} clusters and a \ce{1,3,6,8-Tetra (4-carboxylphenyl)} pyrene ligand encapsulating a bipyramidal Au NP (Au@Nu-1000). These samples were suspended in ethanol and drop-cast onto a carbon-coated \ce{Cu} transmission electron microscopy (TEM) grid. Imaging was performed using a Thermo Fisher Scientific Tecnai Osiris TEM with an acceleration voltage of $200$ kV, a screen current of $50$ pA, and an imaging/scanning dwell time of $3.06/7.96~ \mu$s. Au/Pd NS samples were collected using HAADF-STEM, and MOF complexes were collected using ADF-STEM.

Sample tracking and focusing were performed manually. During RECAST3D imaging, projection alignment was performed by centering the sample, masking the background with an Otsu threshold, and then aligning the projections in chronological order using intensity correlation. Post-processing reconstruction and alignment were performed using the ASTRA toolbox in MATLAB. Before post-processing, the tilt-series was sorted into annular order (lowest to highest, \eg, $-70^\circ$ to $70^\circ$) and projection alignments were performed using intensity correlation.

For comparison, tilt-series were acquired with both IS and GRS. The simulated and acquired tilt-series throughout this work are summarized in Table~\ref{tab:2}. Approximately 70 projections of GRS acquisition were acquired regardless of the proposed termination point for comparison with the standard protocol of IS with a $2^\circ$ increment (71 projections). IS acquisitions were collected with a tilt increment of $2^\circ$, $5^\circ$, $7^\circ$, $10^\circ$, $14^\circ$, $35^\circ$, or $70^\circ$. These are the only integer tilt increments that result in all projections being equally spaced between $\pm 70^\circ$.

\begin{table*}[!b]
\centering

\caption{Collected tilt-series}
\begin{tabular}{c|c|c|c|c|c|c}
\toprule
Type & sample & name & $\beta1: \beta2$ & acquisition type & annular range [$^\circ$] & tilt step [$^\circ$] \\ \midrule
\multirow{8}{*}{simulation} & \multirow{8}{*}{nanocage} & \multirow{2}{*}{NC-1} & 0:0 & GRS & $\pm70$ & \\ 
 &  & & 0:0 & IS & $\pm70$ & 2, 5, 7, 10, 14, 35, 70 \\ 
 &  & \multirow{2}{*}{NC-2} & 0.3:0.03 & GRS & $\pm70$ & \\
 &  & & 0.3:0.03 & IS & $\pm70$ & 2, 5, 7, 10, 14, 35, 70 \\
 &  & \multirow{2}{*}{NC-3} & 0.55:0.055 & GRS & $\pm70$ & \\
 &  & & 0.55:0.055 & IS & $\pm70$ & 2, 5, 7, 10, 14, 35, 70 \\
 &  & \multirow{2}{*}{NC-4} & 0.6:0.06 & GRS & $\pm70$ & \\ 
 &  & & 0.6:0.06 & IS & $\pm70$ & 2, 5, 7, 10, 14, 35, 70 \\ \midrule
\multirow{4}{*}{real} & \multirow{2}{*}{nanostar} & \multirow{2}{*}{Au/Pd NS} & & GRS & $\pm70$ & \\ 
 &  & & & IS & $\pm70$ & 2, 5 \\
 & Au@NU-1000 & Au@NU-1000 & & GRS & $\pm70$ & \\
 & Au/Pd@ZIF-8 & Au/Pd@ZIF-8 & & GRS & $\pm70$ & \\ \bottomrule
\end{tabular}
\label{tab:A1}
\end{table*}

\section{Results}
\label{sec:Results}

\subsection{Simulated evaluation of optimization protocol}
\label{sec:Results:SimEval}

\subsubsection{Incremental scanning.~~} 
\begin{figure*}[!t]
	\centering
	\includegraphics[height=0.71\textheight]{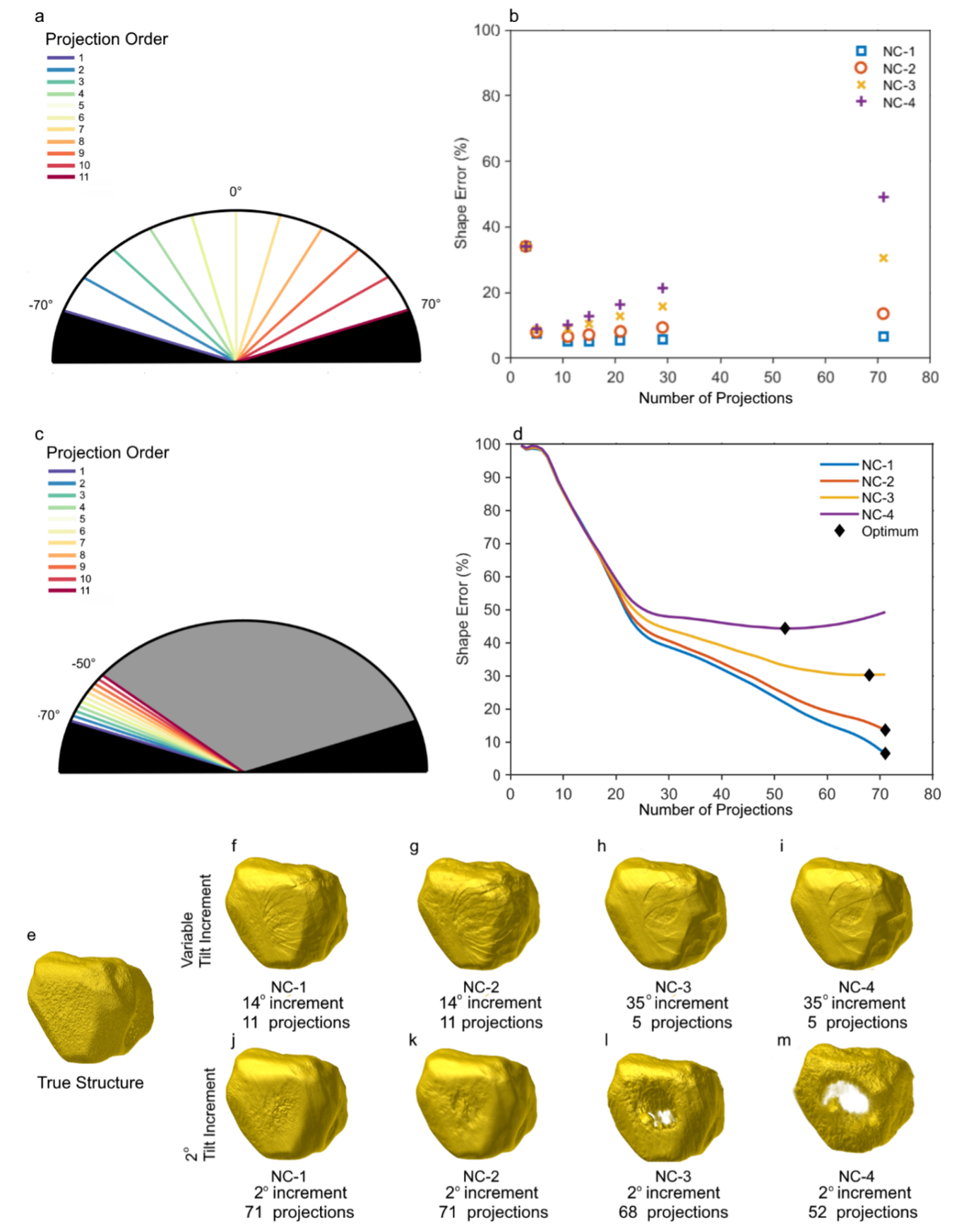}
	\caption{To change the projection number while maintaining a constant annular range, samples NC-1 to NC-4 were collected with a variable tilt increment of  $2^\circ$, $5^\circ$, $7^\circ$, $10^\circ$, $14^\circ$, $35^\circ$, $70^\circ$ (71, 29, 21, 15, 11, 5, 3 projections). An example is shown for the acquisition of 11 projections ($14^\circ$ tilt increment) (a). The  $E_s$ of each acquisition was then determined (b). Alternatively, for samples NC-1 to NC-4, the tilt increment was fixed at a standard $2^\circ$ and more projections were collected while increasing the annular range during a single acquisition. An example is shown for the acquisition of 11 projections (c). The $E_s$ was plotted as a function of  the number of projections (d). The 3D reference structure (e) is shown along with the optimum reconstructions for NC-1 to NC-4 determined with a variable tilt increment (f-i) and $2^\circ$ increment (j-m).}
    \label{fig:5}
\end{figure*}

The traditional approach to optimizing the number of projections is to vary the tilt increment during IS scanning. Therefore, to determine the optimum number of projections using IS for NC-1 to NC-4, we simulated tilt-series with tilt increments of $2^\circ$, $5^\circ$, $7^\circ$, $10^\circ$, $14^\circ$, $35^\circ$, and $70^\circ$ (71, 29, 21, 15, 11, 5, 3 projections, respectively). We measured $E_s$ for each tilt-series by comparing them to a ground truth structure, and the minimum $E_s$ was obtained where the 3D reconstruction most accurately reflected the ground truth. As more beam damage was simulated from NC-1 to NC-4, the minimum $E_s$ value was obtained with fewer projections, but the $E_s$ value at the optimum number of projections increased ($E_s$/Projections: 5.1/11, 6.6/11, 8.7/5, 9.1/5) (Figure~\ref{fig:5}a-b, Figure~B3). While it is possible to estimate the optimum number of projections by simulating seven different tilt-series per sample, this process is infeasible for experimental beam damage analysis.

An alternative method would be to take a single tilt-series using a fixed tilt increment and collect projections until an optimum is obtained. Therefore, we collected a tilt-series for NC-1 to NC-4 for IS with a $2^\circ$ increment while monitoring the $E_s$ as a function of the number of projections (Figure~\ref{fig:5}c). As more beam damage was simulated from NC-1 to NC-4, the $E_s$ optimum was obtained earlier with an increased value (Figure~\ref{fig:5}b), indicating lower quality reconstructions ($E_s$/Projections: 6.6/71, 13.6/71, 30.2/68, 44.3/52). The same trend was observed for variable tilt increments, but the optimum number of projections occurred with far more projections compared to the results displayed in Figure~\ref{fig:5}b. The late optimal number of projections in Figure~\ref{fig:5}d occurs because, until the final projection, each projection is filling a missing wedge in the tilt-series. In contrast, the projections are spread across the entire annular range by taking multiple tilt-series with a variable tilt increment. Therefore, when reducing the number of projections during an IS acquisition, the reduction in beam damage artifacts is counteracted by increased missing wedge artifacts. With NC-4, this is visually apparent. At the $E_s$ optimum of 52 projections, a missing wedge artifact is compensated for when adding new projections, but adding extra projections increases the beam damage artifact (Figure~B2). Through visual inspection, it is apparent that by finding the optimum of multiple tilt-series (Figure~\ref{fig:5}e-i), undersampling artifacts are apparent in samples NC-2 to NC-4. However, by finding the optimum from a single typical acquisition (IS, $2^\circ$)(Figure~\ref{fig:5}j-m), severe beam damage artifacts are apparent in NC-3 and NC-4 (Movie S2).

In summary, current techniques for optimizing the number of projections either require multiple acquisitions or introduce substantial beam damage artifacts while correcting for missing wedge artifacts, limiting their feasibility for beam-sensitive samples. One possible solution to this problem is to use the GRS method, which allows for the determination of the optimum number of projections from a single acquisition. In the following subsection, we describe the GRS method and compare it to the traditional IS method for optimizing the number of projections in beam-sensitive samples.

\subsubsection{Golden ratio scanning.~~} 
\begin{figure*}[!t]
	\centering
	\includegraphics[width=0.8\textwidth]{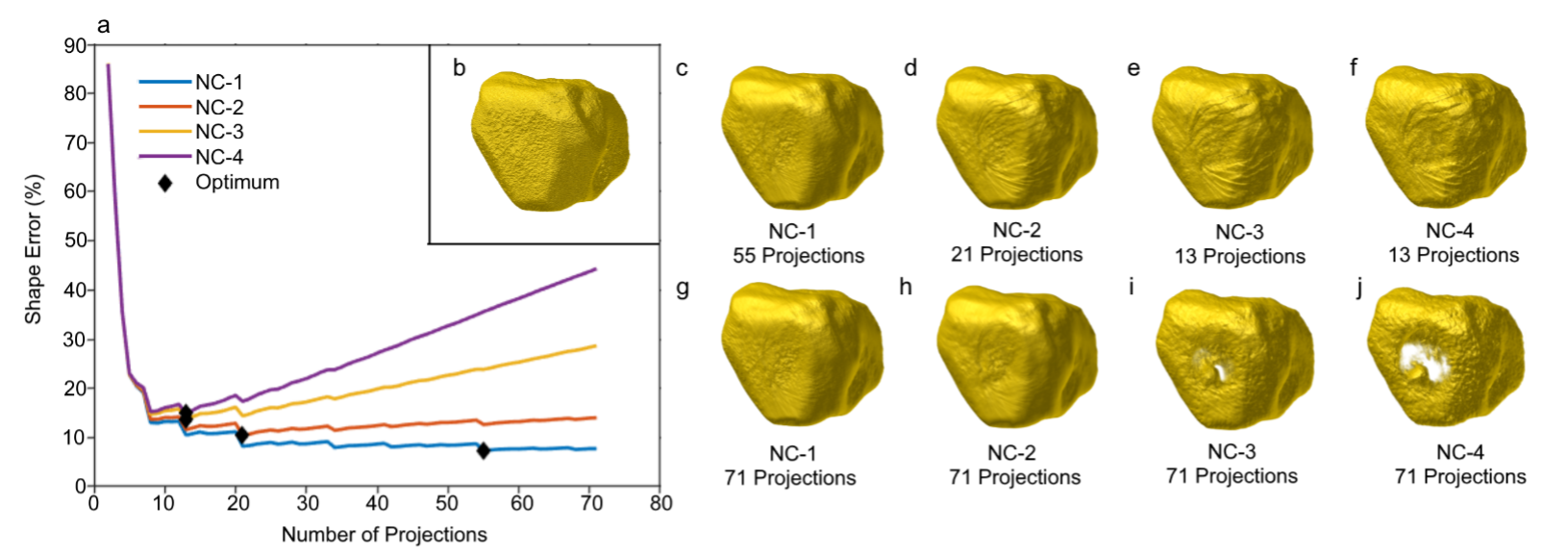}
	\caption{(a) Shape error as a function of the number of projections for NC1-4 nanocages and their determined optimum number of projections with GRS acquisition scheme. (b) Inset: nanocage before beam damage simulation. 3D reconstruction of NC-1-4 with their optimum number of projections (c-f) and 71 projections (g-j).}
    \label{fig:6}
\end{figure*}

In order to determine the optimum number of projections from a single tilt-series, we performed simulations for samples NC-1 to NC-4 according to the golden ratio scanning (GRS) method (Figure~\ref{fig:2}b) and evaluated the reconstruction quality using $E_s$. For each sample, we obtained a local minimum $E_s$ (Figure~\ref{fig:6}a). This minimum represents the optimum between undersampling and beam damage. As more damage was simulated from NC-1 to NC-4, the optimum was found with fewer projections, but the $E_s$ value increased ($E_s$/Projections: 7.4/55, 10.4/21, 13.7/13, 14.9/13). Therefore, reconstructions with fewer projections were favored with increased beam damage simulation because beam damage artifacts outweighed undersampling. However, despite optimizing the number of projections, the overall reconstruction quality worsened as more beam damage was induced.

It is surprising that a local minimum is achieved at 55 projections for NC-1, in which no beam damage was simulated. However, the Es at 71 projections ($7.8\%$) differs from 55 projections by just $0.4\%$. Therefore, the obtained minimum is likely just statistical variance. The obtained optimum number of projections is consistent with the visual inspection of the samples. For NC-1 and NC-2, there is little notable distortion in the reconstruction at the optimum number of projections (Figure~\ref{fig:6}b-d). Slight surface defects were noted in NC-3 and NC-4 (Figure~\ref{fig:6}e-f). However, these were minor compared to the beam damage-induced cavities apparent in NC-3 and NC-4 with 71 projections (Figure~\ref{fig:6}g-j).

When comparing the minimum $E_s$ obtained from NC-1 to NC-4 for IS and GRS (IS($\%$)/GRS($\%$): 6.6/7.4, 13.6/10.4, 30.2/13.7, 44.3/14.9), the $E_s$ value for IS is substantially larger than the same value obtained for GRS for NC-2 to NC-4, indicating that optimization of GRS acquisitions produces a substantially improved reconstruction compared to IS with a standard $2^\circ$ increment for beam-sensitive samples.

\subsubsection{TUOTA.~~} 
\begin{figure*}[!t]
	\centering
	\includegraphics[width=0.8\textwidth]{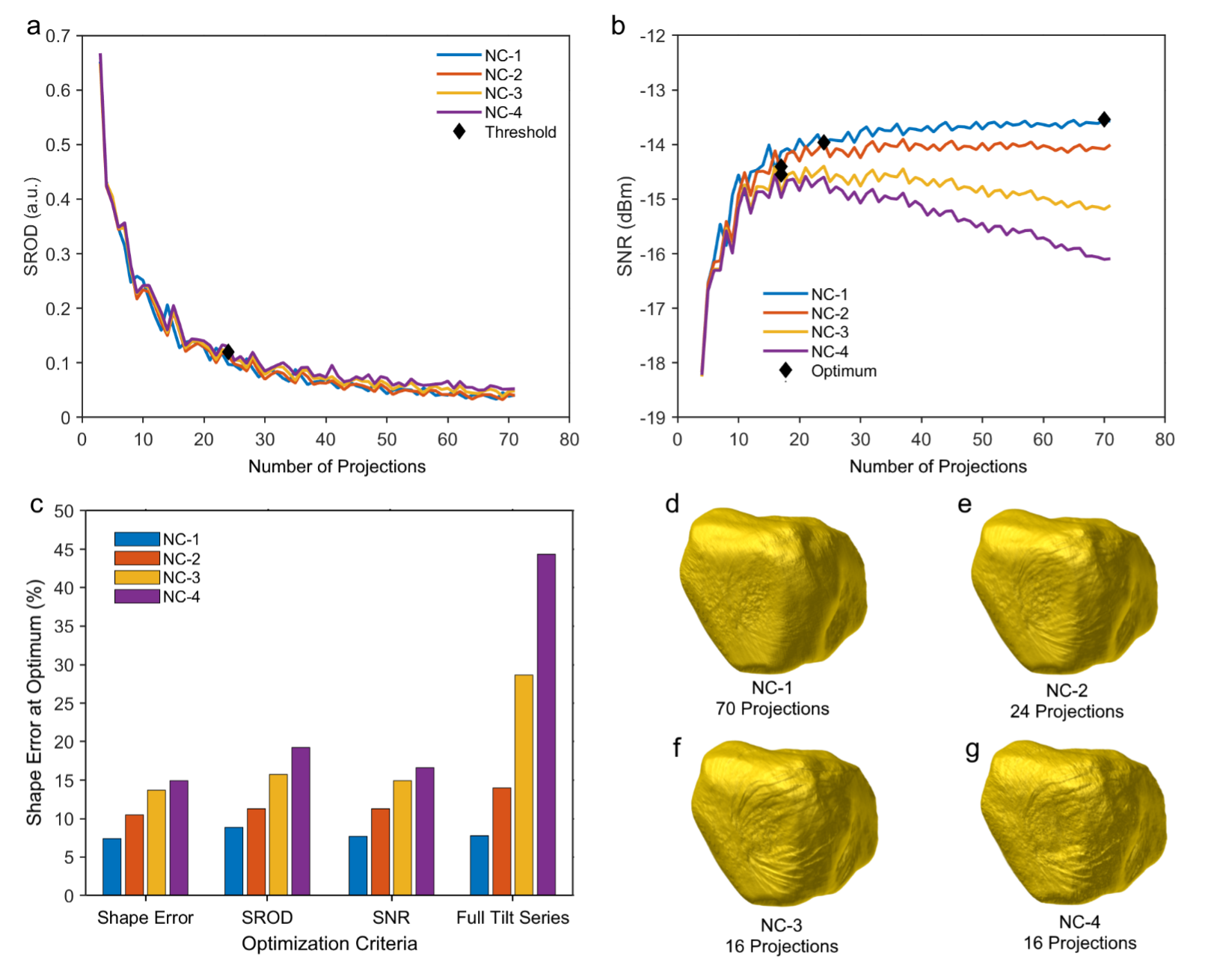}
	\caption{(a) SROD showing the 0.1 threshold and (b) SNR of NC-1 to NC-4 orthoslices determined by TUOTA. (c) Comparison of the $E_s$ for the optimum reconstruction determined from the shape error, SNR, SROD, and complete tilt-series of 71 projection. (d-g) reconstruction of NC-1 to NC-4 with the optimum number of projections determined by SNR.}
    \label{fig:7}  
\end{figure*}

We have previously evaluated reconstruction quality using the metric $E_s$, by comparison to a known ground truth structure. To determine the optimal number of projections in real experiments, it is necessary to do so without prior knowledge of the ground truth. TUOTA has been applied as a promising approach for this purpose, using samples NC-1 to NC-4 and monitoring the SROD and SNR as a function of the number of projections during GRS. The SROD threshold was reached for all NC samples at approximately 24 projections (Figure~\ref{fig:7}a). It is important to note that the number of projections optimized for $E_s$ and SROD identify different properties. The SROD determines the number of projections beyond which additional projections are unlikely to significantly improve reconstruction, while the $E_s$ criteria identify the number of projections that produce the most accurate reconstruction shape. This difference is particularly apparent in the case of NC1-GRS, where the $E_s$ and SROD (using a threshold of 0.1 as described in Section~\ref{sec:Method:TUOTA:Quantification}) identified 55 and 22 projections, respectively. In the absence of damage, the acquisition could be continued indefinitely, but there was no visible change beyond a certain point (Figure~B4). As damage increased from NC-1 to NC-4, the maximum SNR value decreased (Figure~\ref{fig:7}b), occurring at a later projection (SNR(dBm)/Projections: -13.6/70, -14.0/24, -14.4/16, -14.6/16). Thus, with more simulated damage, reconstructions using fewer projections were optimal.
To validate the TUOTA results, we calculated the $E_s$ for the optimal reconstructions determined by TUOTA and compared them to the full tilt series and the optimal reconstruction based on the minimum $E_s$ determined in 3.1.2. While the optimal number of projections determined by $E_s$, SROD, and SNR did vary, the reconstruction quality as determined by $E_s$ remained largely the same (Figure~\ref{fig:7}c, Table~\ref{tab:2}, Movie S3). Visual inspection of the TUOTA-determined optimal reconstructions for NC-1 showed no artifacts. In contrast, NC-2 to NC-4 had a rippled texture due to an artifact at their TUOTA-determined optimal number of projections (Figure~\ref{fig:7}d-g). This is consistent with the optimal reconstructions obtained by $E_s$ (Figure~\ref{fig:6}c-f). These results demonstrate that, for simulated nanocages, TUOTA can accurately determine the optimal number of projections, comparable to a reference ground truth structure.

\begin{table}[!b]
    \centering
    \caption{Shape error $E_s$ and number of projections (NPs) of GRS acquired reconstructions using various optimization criteria, along with full tilt-series}
    \begin{tabular}{ccccccccc}
    \toprule
    \multirow{2}{*}{Sample} & \multicolumn{2}{c}{$E_s$} & \multicolumn{2}{c}{SROD} & \multicolumn{2}{c}{SNR} & \multicolumn{2}{c}{full} \\
    & NPs & $E_s$ & NPs & $E_s$ & NPs & $E_s$ & NPs & $E_s$ \\ \midrule
    NC-1 & 55 & 7.4 & 22 & 8.8 & 70 & 7.7 & 71 & 7.7 \\
    NC-2 & 24 & 10.4 & 22 & 11.3 & 24 & 11.3 & 71 & 14.0 \\
    NC-3 & 13 & 13.7 & 24 & 15.7 & 16 & 14.9 & 71 & 28.7 \\
    NC-4 & 13 & 14.9 & 24 & 19.2 & 16 & 16.6 & 71 & 44.3 \\ \bottomrule
    \end{tabular}
    \label{tab:2}
\end{table}

\subsection{Experimental validation}
\label{sec:Results:ExpVal}

\subsubsection{Au/Pd nanostar}
As mentioned earlier, a challenge in optimizing tilt undersampling for beam-sensitive samples is the lack of knowledge of the material's true volume. Simulated experiments address this challenge by allowing the comparison of the reconstruction to a simulated "true" reference volume. However, challenges with focusing and aligning projections are not present in simulated data. For samples resistant to beam damage, it can be assumed that the reconstruction obtained with standard ET is a reasonable representation of the sample's true volume. As such, the reconstruction quality can be determined using $E_s$, where the reference is a non-beam-sensitive sample acquired with standard ET.

\begin{figure*}[!t]
	\centering
	\includegraphics[width=0.8\textwidth]{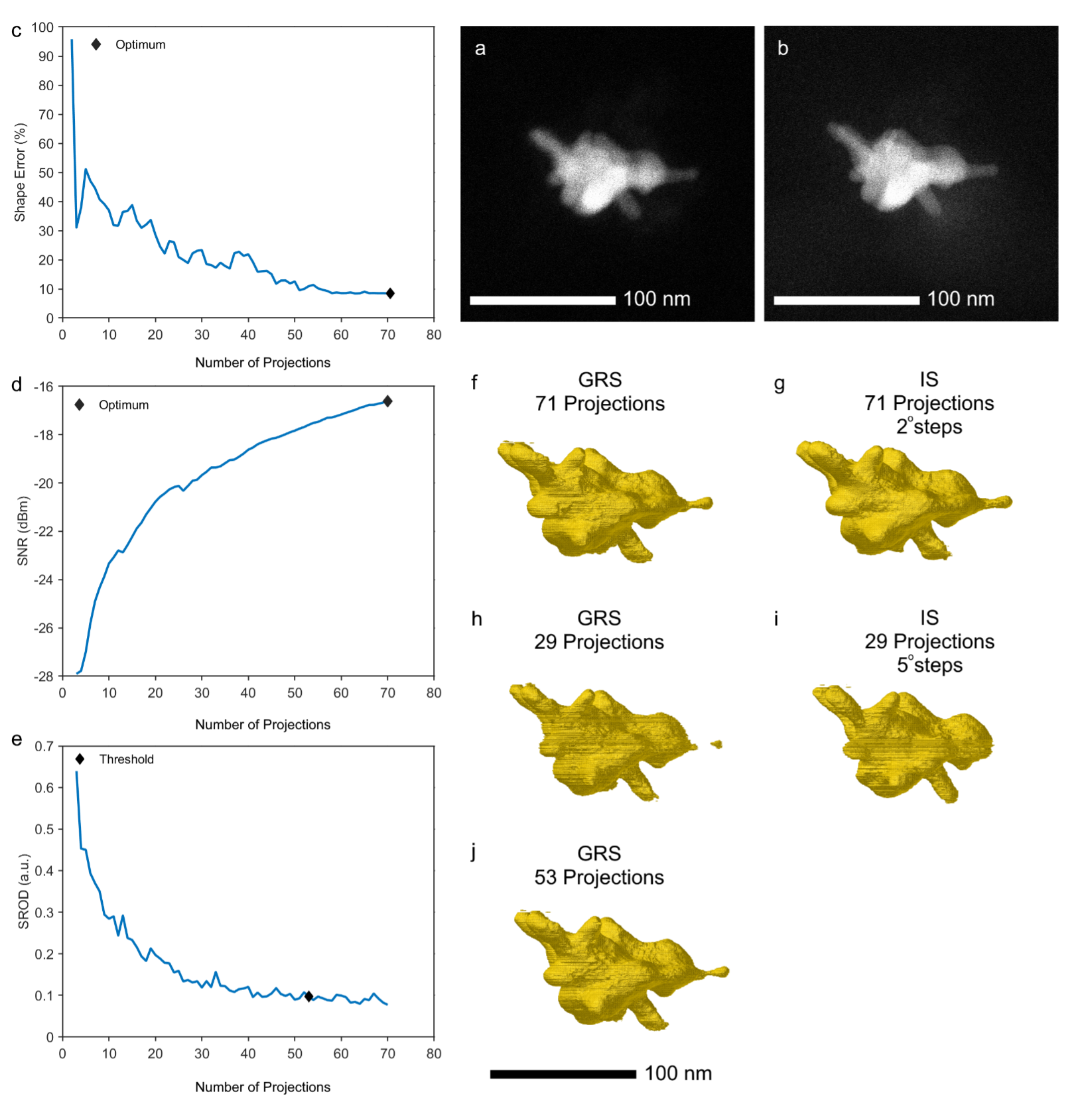}
	\caption{$0^\circ$ projections of Au/Pd NS before (a) and after (b) collection of three tilt-series (IS $2^\circ$, $5^\circ$ increment and GRS 71 projections). The shape error as a function of the number of projections (c) for GRS reconstructions was determined by comparison to the IS reconstruction with a $2^\circ$ increment. The SNR (d) and SROD (e) were determined in real-time using TUOTA. The Au/Pd NS was reconstructed with 71 (f-g) and 29 projections (h-i) for GRS and IS. The GRS tilt-series was also reconstructed with the optimum number of projections determined from the SROD threshold (j). }
    \label{fig:8}
\end{figure*}

To validate TUOTA using experimental data, an Au/Pd nanostar was used as a beam damage-resistant sample. Three acquisitions were performed sequentially on the same sample: a GRS acquisition (71 projections) and an IS acquisition with a $2^\circ$ (71 projections) and $5^\circ$ (29 projections) increment. A $0^\circ$ projection was acquired before and after all collections were completed. Visual inspection of these images showed no obvious signs of beam damage (Figure~\ref{fig:8}a-b).

For GRS reconstructions, the $E_s$ was measured as a function of the number of projections by comparing it to a reference sample collected with IS using a $2^\circ$ increment. Similar to the simulated results for NC-1, with a non-deforming sample, the $E_s$ tends to decrease as more projections are added, but a local minimum is never achieved and the $E_s$ plateaus around 58 projections ($E_s = 8.52\%$) (Figure~\ref{fig:8}c). When adding further projections, the $E_s$ reduces insubstantially to $8.47\%$ (71 projections), indicating a limited improvement to the reconstruction. At 58 and 71 projections, the 3D structure is visually indistinguishable (Figure~B5). When applying TUOTA, the SNR increases but plateaus as more projections are added (Figure~\ref{fig:8}d). The maximum SNR (-16.6 dBm) is obtained when the full tilt-series is collected, indicating there is no beam damage reducing the signal quality. As for the SROD, the threshold is reached at 53 projections, indicating a termination point where further projections are unnecessary (Figure~\ref{fig:8}e). At 53 projections, the $E_s$ varies from the minimum $E_s$ by only $2.43\%$ (Table~\ref{tab:3}), indicating a limited difference between the reconstruction with 71 and 53 projections.

Through visual inspection of the sample, no artifacts are apparent when comparing the full GRS reconstruction with the IS reconstruction with 2◦ steps (Figure~\ref{fig:8}f-g). However, when the number of projections decreases to 29 projections, artifacts are apparent for both GRS and IS (Figure 8h-i, Movie S4). As mentioned earlier, reconstruction convergence was identified at 53 projections using TUOTA. Hence, the GRS tilt-series with 53 projections was reconstructed (Figure~\ref{fig:8}j). No noticeable difference between this reconstruction and the reference structure could be seen (Table~\ref{tab:3}).

Overall, as expected for Au/Pd nanoparticles, beam damage could not be identified either through visual inspection of the sample or analysis. An optimum number of projections between 29 and 71 projections was obtained through IS acquisition. Using TUOTA, the optimum number of projections was narrowed to 53 projections during a single acquisition. This demonstrates the effectiveness of TUOTA in determining an optimal number of projections without the need for a ground truth reference structure, even when applied to beam-resistant samples.

\begin{table}[!t]
    \centering
    \caption{$E_s$ and number of projections of a Au/Pd nanostar acquired by GRS terminated using various optimization criteria.}
    \begin{tabular}{r | cc}
    \toprule
    Optimization Criteria & number of projections & $E_s(\%)$ \\ \midrule
    Minimum $E_s$ & 71 & 8.47 \\
    SROD & 53 & 10.9 \\
    SNR & 71 & 8.47 \\
    Full & 71 & 8.47 \\ \bottomrule
    \end{tabular}
    \label{tab:3}
\end{table}

\subsubsection{NP@MOF composite}
The technique TUOTA was applied to two MOF composites: Au@NU-1000 and Au/Pd@ZIF-8, which are known to undergo significant changes in shape and crystallinity when exposed to a beam. \cite{liu2020bulk,aulakh2019direct,rosler2014encapsulation} Using a GRS technique, the degradation and contamination of the samples were observed by comparing the first and last projections collected (Figure~B6). The crystal facets became less defined and a large, blurry ring appeared around the sample, indicating the presence of carbon contamination.

\begin{figure*}[!t]
	\centering
	\includegraphics[width=\textwidth]{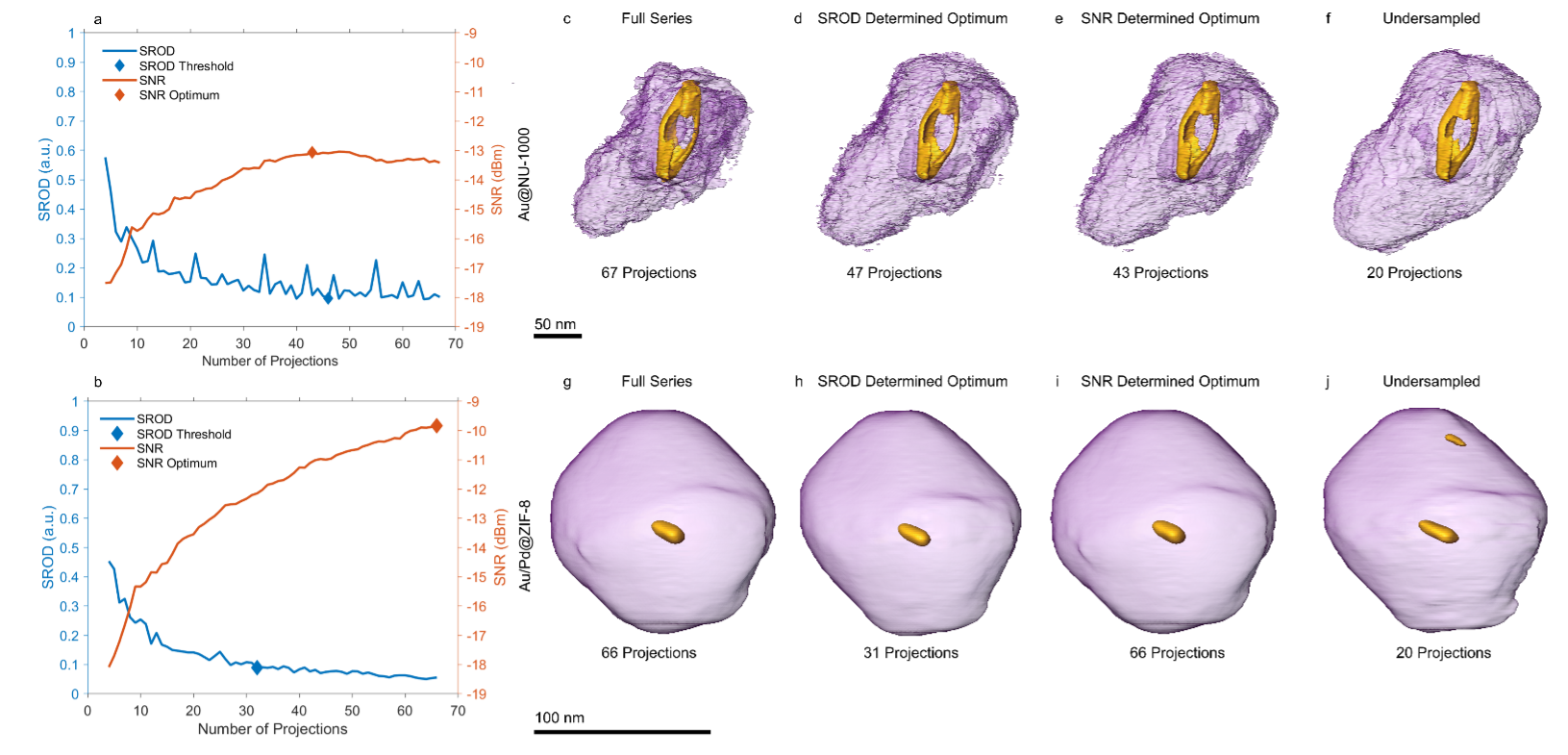}
	\caption{SNR (a) and SROD (b) as a function of the number of projections for Au@NU-1000 and Au/Pd@ZIF-8. 3D reconstructions were acquired with the complete tilt-series, SNR optimum, SROD threshold, and 20 projections (left to right) for Au@NU-1000 (c-f) and Au/Pd@ZIF-8 (g-j).}
    \label{fig:9}
\end{figure*}

The SNR was also analyzed (Figure~\ref{fig:9}a). The Au/Pd@ZIF-8 sample had a higher maximum SNR (SNR of -9.84 dBm with 66 projections) than the Au@NU-1000 sample (SNR of -13.1 dBm with 43 projections). Additionally, the maximum SNR was achieved at the end of the tilt-series for Au/Pd@ZIF-8, while it occurred at a local maximum for Au@NU-1000. This suggests that the Au@NU-1000 sample is more sensitive to beam-induced deformation, consistent with the SROD results. The SROD threshold for Au/Pd@ZIF-8 was obtained at 31 projections, indicating that while the SNR improved with additional projections, the reconstruction showed minimal change past 31 projections. In contrast, the SROD threshold for Au@NU-1000 was achieved later, at 43 projections, and was generally higher and less consistent than that of Au/Pd@ZIF-8, indicating difficulty in converging to a consistent reconstruction due to additional noise (Figure~\ref{fig:9}b).

Visual inspection of the samples supports the findings of the TUOTA analysis. For Au@NU-1000, the NU-1000 shell displayed substantially more surface detail at the SNR optimum compared to the sample particle undersampled with 20 projections. Using the full tilt-series, the MOF shell had significantly shrunk, indicating continued deformation (Figure~\ref{fig:9}c-f, Movie S5). In the case of Au/Pd@ZIF-8, little difference was observed between the full tilt-series, SNR optimum, and SROD optimum. Undersampling with 20 projections resulted in a reconstruction in which the Au/Pd nanoparticle could not be properly segmented (Figure~\ref{fig:9}g-j, Movie S6).

Overall, the results of this study suggest that TUOTA can be used to determine the optimal acquisition point for MOF samples to prevent beam damage. For the NU-1000 sample, acquisition should be terminated at 43 projections. For Au/Pd@ZIF-8, beam damage is evident, but it has a limited impact on reconstruction quality, and acquisition can be terminated after 31 projections to obtain good results.

\section{Discussion}

\subsection{Tilt scheme}
In our study, we found that using RECAST3D and GRS scanning in both experimental and simulated cases resulted in reconstructions that were comparable to or better than those obtained with IS using a standard tilt increment of 2°. However, when the number of projections in the IS scan was optimized to be similar to that of the GRS scan, there was a slight decrease in the reconstruction quality measured with Es. One possible explanation for this finding is that GRS tends to sample almost the entire range of tilts, or “annular range," but falls short of fully covering it. For instance, in a tilt range of $\pm 70^\circ$ ($140^\circ$ total), the first ten projections of a GRS scan cover about $119.6^\circ$, increasing to $127.4^\circ$ (20 projections) and $132.2^\circ$ (30 projections). In contrast, IS always samples the full annular range regardless of the tilt increment used. It is noted that whilst optimized IS may provide slight improvement on the reconstruction quality over GRS, optimizing the number of projections in an IS scan is not feasible for beam-sensitive samples. Furthermore, even if optimization were possible, the need to acquire multiple tilt series would make the process time-consuming. As an alternative, it may be beneficial to consider a two-step approach in which the optimal number of projections is first determined using GRS, followed by the acquisition of a second tilt series using IS with a tilt increment that approximates the optimal number of projections found using GRS. This approach could potentially lead to a slightly improved reconstruction while also reducing the beam exposure time due to the tracking and refocusing steps required in GRS imaging. It is worth noting that in our study, GRS tracking and focusing were performed manually, but automated tracking could significantly reduce the beam exposure time in GRS imaging.

\subsection{Software architecture}
Most of TUOTA is implemented using RECAST3D, as described in previous studies. \cite{buurlage2018real, buurlage2019real} However, there are additional constraints for quantification that require modifications to RECAST3D. In particular, the orthoslices at $N-1$ projections must have the same orientation and tilt axis as the orthoslices with $N$ projections. While RECAST3D allows these parameters to be adjusted in real-time, doing so would invalidate the quantification results of TUOTA and prevent the user from visually inspecting other regions of the sample or correcting the tilt-axis alignment. Additionally, the default orthoslices selected by RECAST3D ($xy$, $xz$, and $yz$ slices passing through the origin) may not be representative slices of the entire volume. For example, in the case of an 8-dendrite nanostar, these slices could go through the center and miss every dendrite, resulting in a large region of the sample being outside the inspected area (Figure~B7). To address this issue, it is possible to visually inspect the sample and adjust the orthoslice selection by rotating the $xy$ and $xz$ planes $45^\circ$, resulting in a more representative slice of the volume.

\subsection{Alignment}
Projection alignment is a major challenge during TUOTA. In previous studies, projection and tilt axis alignment have been performed in real-time using RECAST3D.\cite{vanrompay2020real} However, when applying TUOTA to beam-sensitive samples, there are some additional challenges to consider. Firstly, intensity cross-correlation can result in poorly aligned projections due to the inclusion of other features in the images, such as beam-damaged regions of the carbon mesh, other particles, or the grid. To address this issue, we use watershed segmentation to identify the largest particle in the image and mask out everything else.

The second challenge is that GRS typically has large annular distances between projections, which can lead to inaccuracies during cross-correlation. For example, the second projection ($-37.0^\circ$) and the third projection ($49.6^\circ$) are separated by $86.5^\circ$. To address this issue, we index images by both angle and chronology and align each projection to the closest projection by angle, rather than aligning to the previously collected projection. Overall, our method for addressing these challenges has been successful in ensuring accurate projection alignment during TUOTA of beam-sensitive samples.

\section{Conclusions}
In conclusion, we have developed a novel protocol for optimizing tilt undersampling during a single acquisition using GRS and RECAST3D. Our simulations have demonstrated that reconstructions of beam-sensitive samples optimized using this method have higher fidelity with the pre-damaged sample than reconstructions using standard incremental acquisition. We have validated our approach through simulations and experimental 3D imaging of Au/Pd nanostars and applied it to the characterization of highly sensitive NP@MOF complexes. Our approach, which is based on golden ratio acquisition and quasi-real-time reconstruction, provides an effective solution for balancing undersampling, beam damage, and reconstruction quality on a sample-by-sample basis. While similar results can be achieved with undersampling optimization of IS, our method is far more efficient and less time-consuming.

Future work may involve further optimization and testing of the TUOTA protocol on a wider range of beam-sensitive samples and comparing its performance to other acquisition schemes. Additionally, exploring the use of more advanced reconstruction algorithms in conjunction with TUOTA could potentially lead to even higher-quality reconstructions.


\section*{Funding Statement}
This project received funding received from the European Union’s Horizon 2020 research and innovation programme under grant agreement no. 860942 and from the European Research Council under the ERC Consolidator Grant no. 815128 REALNANO.

\section*{Conflicts of interest}
The authors declare no financial interests/personal relationships that may be considered potential competing interests.

\section*{Acknowledgements}
The authors would like to acknowledge the financial support received from the European Union's Horizon 2020 research and innovation program through grant agreement no. 860942 - HEATNMOF. S.B. and A.A.K. also acknowledge support from the European Research Council (ERC Consolidator Grant no. 815128 REALNANO). The authors are grateful for the assistance provided by Armand Béché, Lars Riekehr, and Daniel Arenas Esteban at the EMAT of the University of Antwerp, including training on and use of the TEM, as well as assistance with 3D visualization and rendering of nanomaterials. The authors also acknowledge the contribution of samples from Pablo del Pino and his research group at the University of Santiago de Compostela for use in the characterizations presented in this study.

\balance



\newpage
\appendix 
\counterwithin{figure}{section}
\counterwithin{equation}{section}
\counterwithin{table}{section}

\section{Methods}
\subsection{RECAST3D}

\begin{figure}[!b]
	\centering
	\includegraphics[width=\columnwidth]{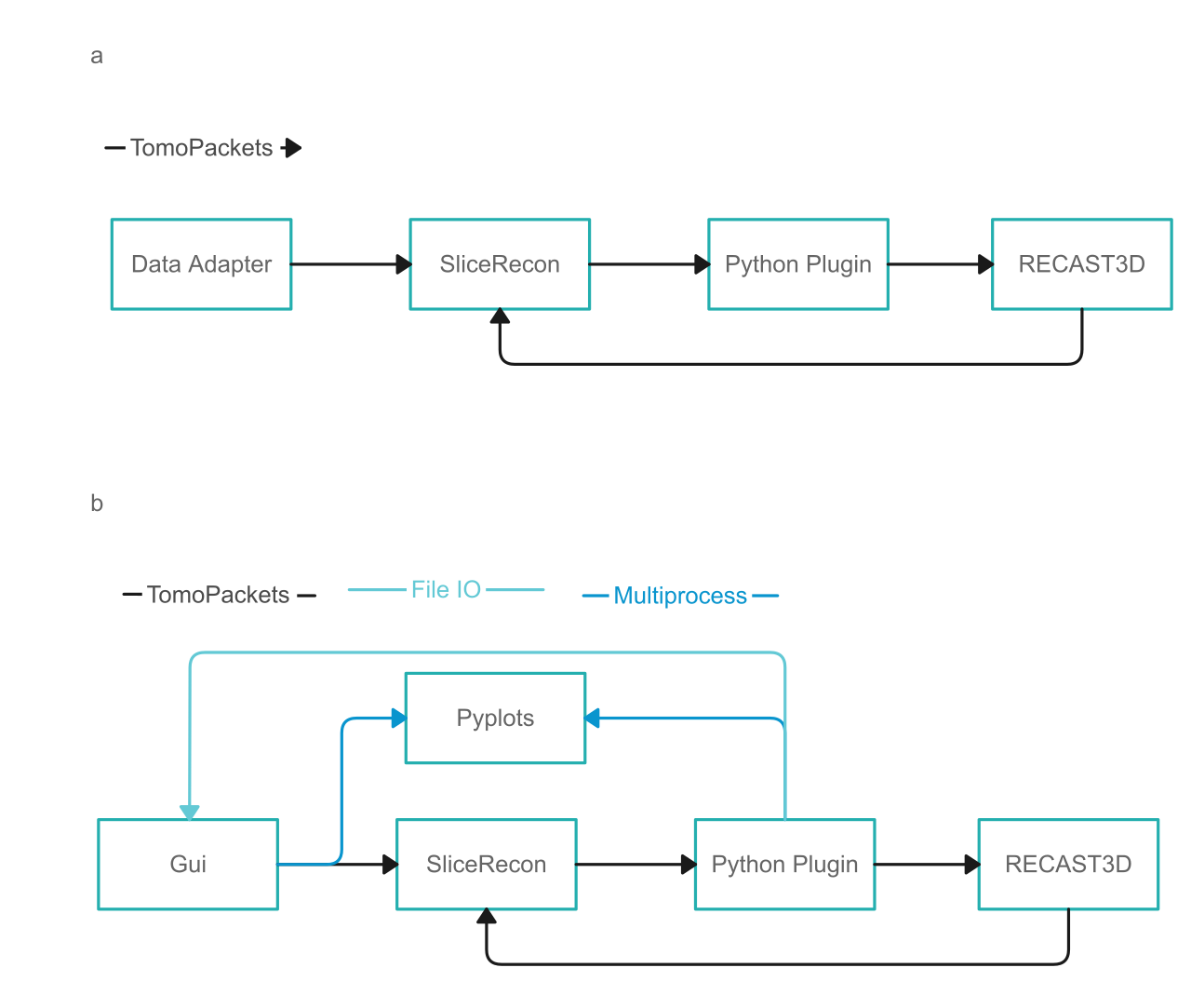}
	\caption{Illustration of RECAST3D workflow prior (a) and post modification (b). The primary modifications involved the substitution of the data adapter with a GUI, the implementation of a plotting library using multiprocessing for visualizing analysis and alignment data from the GUI and plug-in, and the introduction of a file input-output (IO)-based method for communication between the GUI and plug-in through the reading and writing of settings. These modifications aimed to enhance the functionality and usability of the RECAST3D workflow.}
    \label{fig:s1}
\end{figure}

RECAST3D is a tool for real-time, quasi-3D reconstruction and visualization. When used at a microscope, it enables real-time viewing of three reconstructed orthoslices directly from collected images. The full code and instructions for operation and installation can be found at \url{https://github.com/cicwi/RECAST3D}. In this section, we describe the implementation and modifications made to RECAST3D for the purposes of this work.

RECAST3D consists of three libraries: the main library for visualizing orthoslice data, SliceRecon for performing reconstructions from projection data, and Tomopackets for transporting data between layers via packets (Figure~\ref{fig:s1}a). Buurlage et al. developed a framework for adding custom python scripts to increase the flexibility of this architecture.\cite{buurlage2018real} One such script is the data adapter, which allows users to import saved projection data, create projection geometries, and perform alignments before orthoslice reconstruction by SliceRecon. Another potential add-on script is a post-reconstruction plugin, in which we developed \url{beam_analysis.py} to calculate quantitative indicators (SROD and SNR) as described in the main text.

To use RECAST3D for beam damage analysis (Figure~\ref{fig:s1}b), we made several changes to address two crucial issues in the RECAST3D workflow. The first issue was the difficulty of performing alignments during real-time analysis. To address this, we added the Python Pyplots package for animating tilt-series images and inspecting alignment quality.\cite{hunter2007matplotlib}. We also implemented Python multiprocessing to view the tilt-series images, preventing the reconstruction from slowing down during the viewing process.\cite{vanrossum} The tilt-series images were simultaneously piped to SliceRecon and Pyplots, and the Pyplots interface was also used to plot the SROD and SNR data from the plugin.

The second issue with RECAST3D for beam damage analysis was that the plugin could not access the slice orientation. This is because the plugin calculates the SNR and SROD by comparing the current orthoslices to the previous set of orthoslices. However, the user can change the slice orientation at will, which would result in comparing two sets of slices with different positions/orientations. To address this issue, we modified Tomopackets to pass the orientation to the plugin. In the case of a change in orientation, a request is sent to the data adapter to restart the reconstruction and SROD/SNR analysis. Finally, for convenience, we replaced the data adapter script with a graphical user interface.

\subsection{Simulations}
\begin{figure}[!b]
	\centering
	\includegraphics[width=\columnwidth]{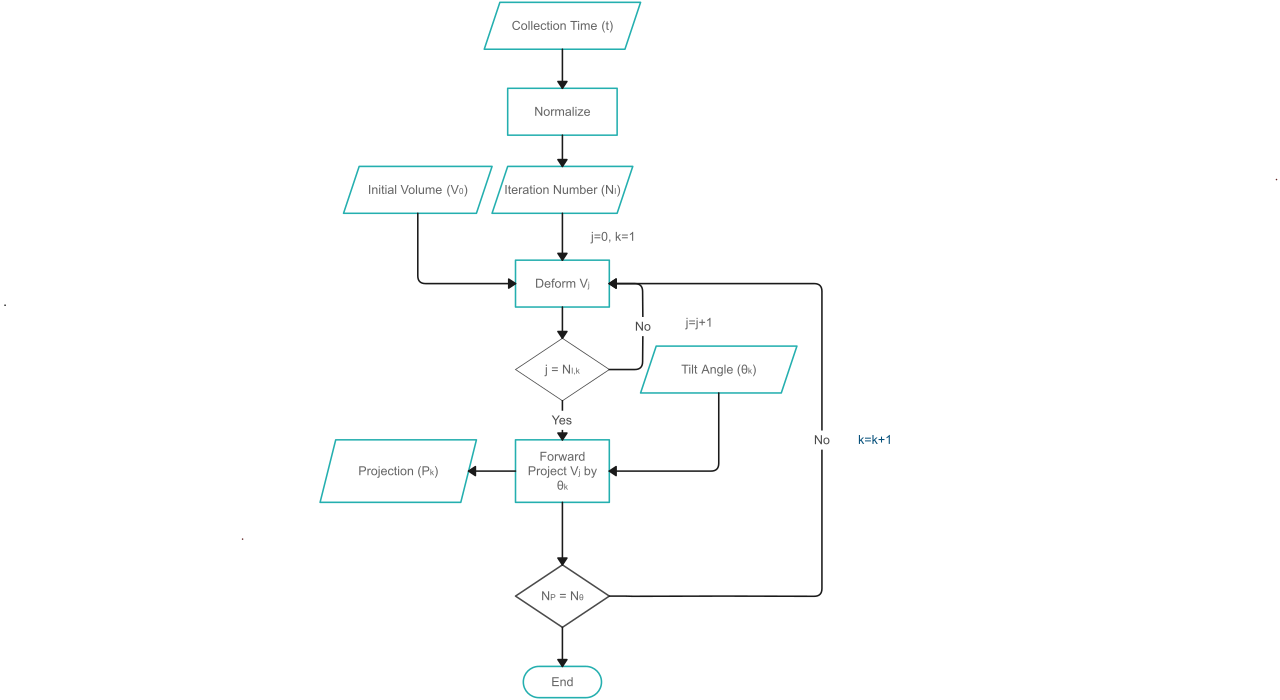}
	\caption{Simulations are performed by iteratively deforming an initial volume for a set number of iterations. An artificial tilt-series is generated from the simulations by iteratively forward projecting the deformed volumes by the angle of collection. The process lasts until a projection is acquired for every angle in the tilt-series.}
    \label{fig:s2}
\end{figure}

Simulations were conducted using an iterative routine as depicted in Figure~\ref{fig:s2}. The input parameters included an initial 3D volume ($V_0$), two deformation parameters ($\beta_1, \beta_2$), a set of collection times ($t$), and tilt angles ($\theta$). The collection times correspond to the time elapsed between the collection of each projection in the tilt series. For instance, a set of $[1, 2.5]$ indicates a tilt series with two projections, the first of which was collected at a time of 1 arbitrary unit, and the second was collected 2.5 units after the first. The collection times were normalized by the minimum collection time and rounded to the nearest integer to determine an iteration number ($N_I$) according to 
\begin{equation}
	N_I = \nint*{\frac{t}{\min (t) } } .
\end{equation}

A deformation function ($f_{\text{def}}$), described in the subsequent subsection, was applied iteratively to the volume from $j = 0$ to the sum of $N_I$, as shown below:
\begin{equation}
	V_j = f_{\text{def}} \left( V_{j-1}, \beta_1, \beta_2 \right)
\end{equation}

The set $N_I$ defines the number of deformation iterations applied between projections. For example, the initial volume is deformed four times for the set $N_I = [1, 3]$, as the first projection is collected after 1 iteration of deformation, while the second projection is collected three iterations after the first. It is worth noting that the projections are obtained by forward projecting the deformed volume $V_j$ at the defined tilt angle $\theta_j$, as described in the following:    
\begin{equation}
	P_j = \text{FP}\left( V_j, \theta_j \right),
\end{equation}
where FP represents forward projection operation.

Two assumptions were made about the acquisition times for the sake of convenience. Firstly, the time taken to collect each projection in a tilt series is the same, \eg , $t = [1, 2, 3, \dots]$. Secondly, the collection time for each projection during GRS and IS acquisition is the same. Specifically, only 1 iteration of deformation per projection was considered in all simulation experiments, regardless of the acquisition angle or collection method.

\subsubsection{Beam Damage Deformation.~~}
In order to simulate beam damage, we employ a two-step deformation process that is parameterized by $\beta_1$ and $\beta_2$. The first step involves applying elastic deformation to the sample through the use of a voxel mask (M) with random values ranging from -1 to 1. A gaussian filter is then applied to smooth the values of the mask based on the intensities of adjacent voxels, using a standard deviation of 10, which indicates that the random array is smoothed with the nearest 30 voxels for each voxel. The voxel mask is then scaled by $\beta_1$, and the voxels of the sample (V) are transformed according to the mask M using elementwise matrix multiplication, \ie,  
\begin{equation}
	V_i = M \odot V_{i-1},
\end{equation}
where $\odot$ represents elementwise matrix multiplication. In the second step, we simulate knock-on damage. We first determine the number of non-zero neighboring voxels (NN) for every non-zero voxel in the sample. We then compute the probability P for each voxel based on NN and the deformation parameter $\beta_2$ using 
\begin{equation}
	P = \begin{cases}
    \beta_2^{NN/3} & \quad \text{if } NN < 27\\
    0  & \quad NN = 27
  \end{cases} .
\end{equation}
If a random number between 0 and 1 is greater than probability P for a given voxel, the intensity of that voxel is set to 0. 
\subsection{Simulated and Experimental Samples}
The settings are tabulated in Table~\ref{tab:A2}. 

\begin{table*}[!htb]
\centering

\caption{Collected tilt-series}
\begin{tabular}{c|c|c|c|c|c|c}
\toprule
Type & sample & name & $\beta1: \beta2$ & acquisition type & annular range [$^\circ$] & tilt step [$^\circ$] \\ \midrule
\multirow{8}{*}{simulation} & \multirow{8}{*}{nanocage} & \multirow{2}{*}{NC-1} & 0:0 & GRS & $\pm70$ & \\ 
 &  & & 0:0 & IS & $\pm70$ & 2, 5, 7, 10, 14, 35, 70 \\ 
 &  & \multirow{2}{*}{NC-2} & 0.3:0.03 & GRS & $\pm70$ & \\
 &  & & 0.3:0.03 & IS & $\pm70$ & 2, 5, 7, 10, 14, 35, 70 \\
 &  & \multirow{2}{*}{NC-3} & 0.55:0.055 & GRS & $\pm70$ & \\
 &  & & 0.55:0.055 & IS & $\pm70$ & 2, 5, 7, 10, 14, 35, 70 \\
 &  & \multirow{2}{*}{NC-4} & 0.6:0.06 & GRS & $\pm70$ & \\ 
 &  & & 0.6:0.06 & IS & $\pm70$ & 2, 5, 7, 10, 14, 35, 70 \\ \midrule
\multirow{4}{*}{real} & \multirow{2}{*}{nanostar} & \multirow{2}{*}{Au/Pd NS} & & GRS & $\pm70$ & \\ 
 &  & & & IS & $\pm70$ & 2, 5 \\
 & Au@NU-1000 & Au@NU-1000 & & GRS & $\pm70$ & \\
 & Au/Pd@ZIF-8 & Au/Pd@ZIF-8 & & GRS & $\pm70$ & \\ \bottomrule
\end{tabular}
\label{tab:A2}
\end{table*}

\newpage
\section{Experimental Data}

\begin{figure}[!htb]
	\centering
	\includegraphics[width=\columnwidth]{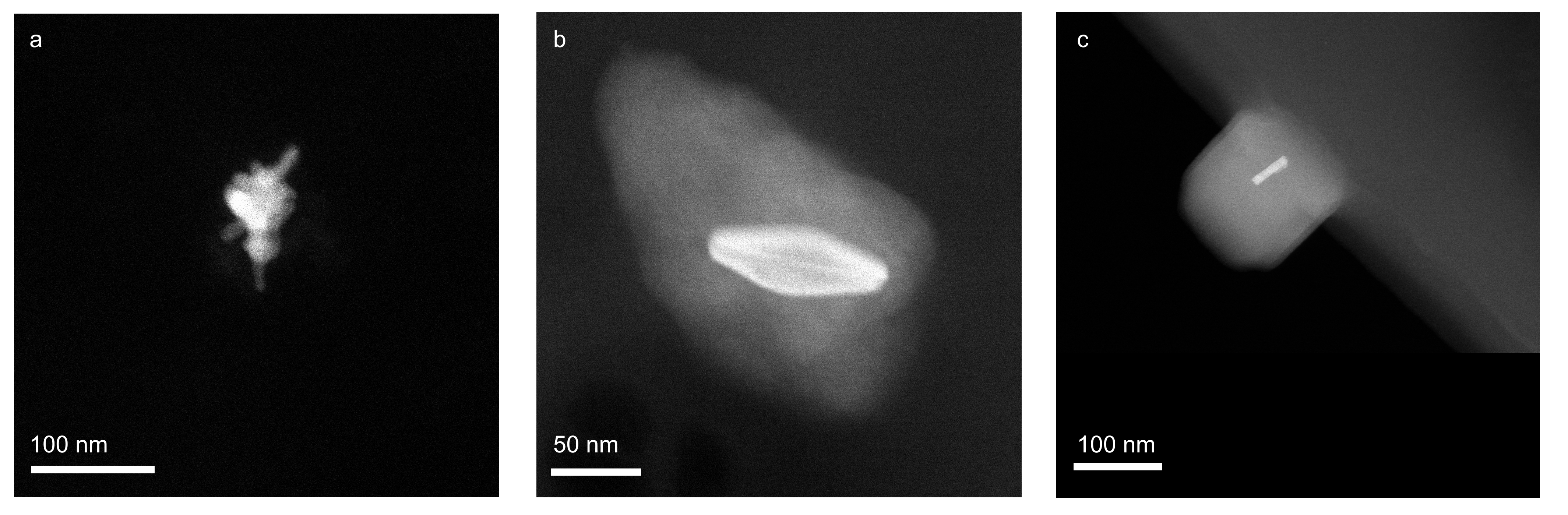}
	\caption{Projection of a Au/Pd Nanostars collected with HAADF (a) and Au@NU-1000 (b) and Au/Pd@ZIF-8 (c) NP@MOF complexes collected with ADF.}
    \label{fig:s3}
\end{figure}

\begin{figure}[!htb]
	\centering
	\includegraphics[width=0.6\columnwidth]{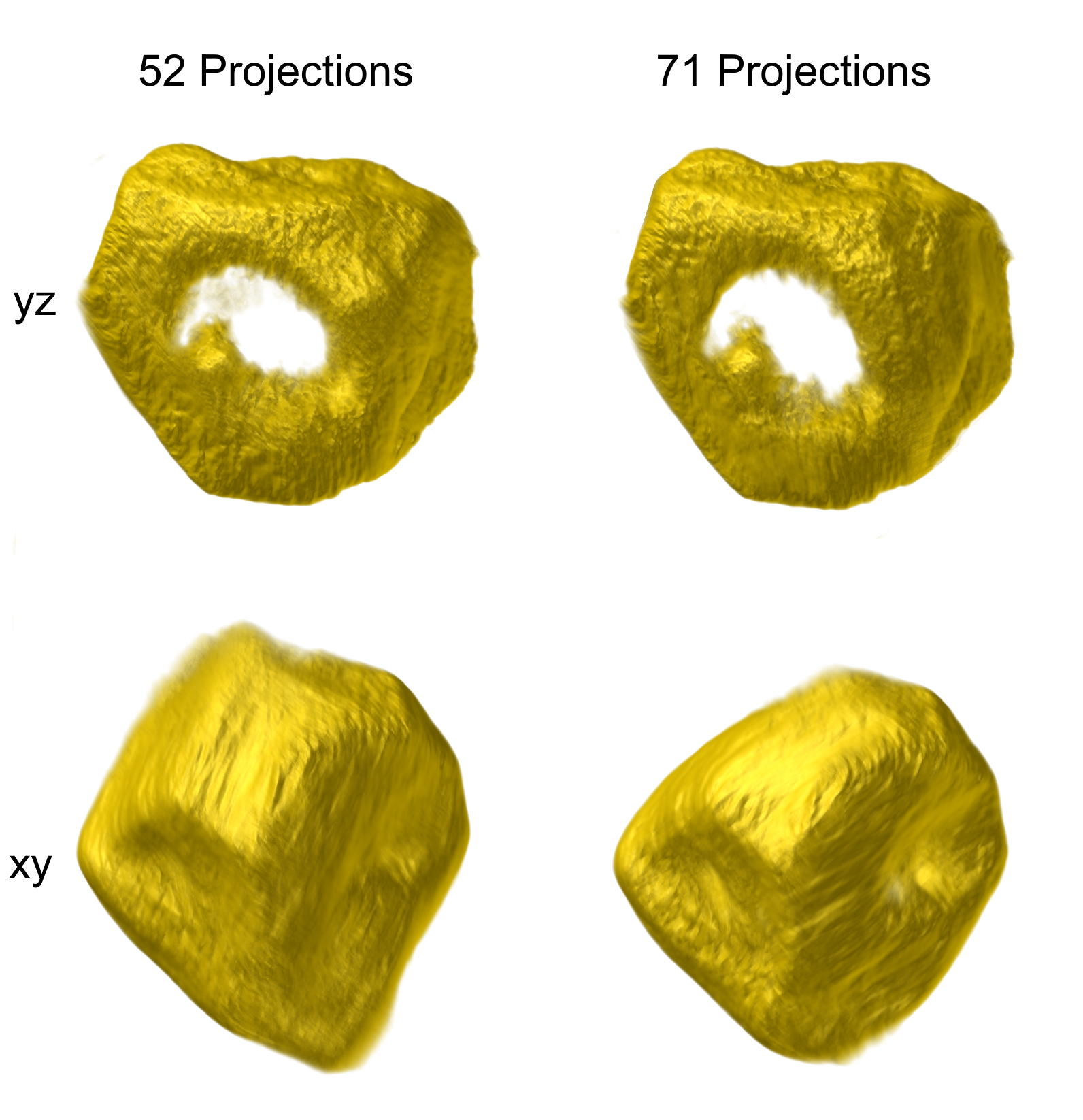}
	\caption{3D reconstruction of NS-4 acquired with IS (2° increment) terminated with 52 or 71 projections shown along the yz and xz plane. As more projections are collected from 52 to 71 projections, missing wedge artefacts in the xy plane are corrected but at a trade-off of increased beam damage artefacts visible in the yz plane.}
    \label{fig:s4}
\end{figure}

\begin{figure*}[!htb]
	\centering
	\includegraphics[width=\textwidth]{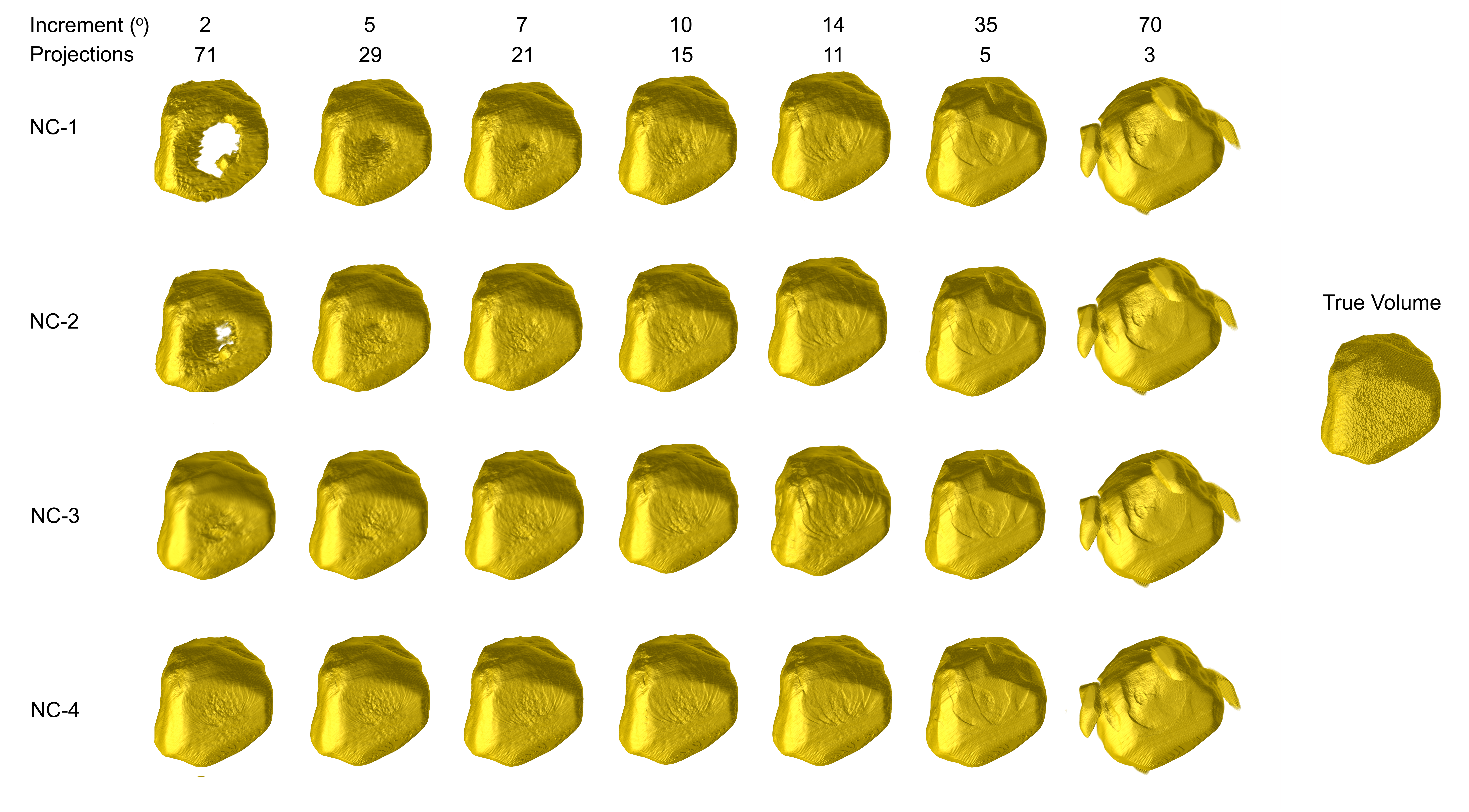}
	\caption{3D reconstruction of NC-1 to NC-4 collected using IS acquisition with a variable tilt increment.}
    \label{fig:s5}
\end{figure*}

\begin{figure*}[!htb]
	\centering
	\includegraphics[width=0.6\textwidth]{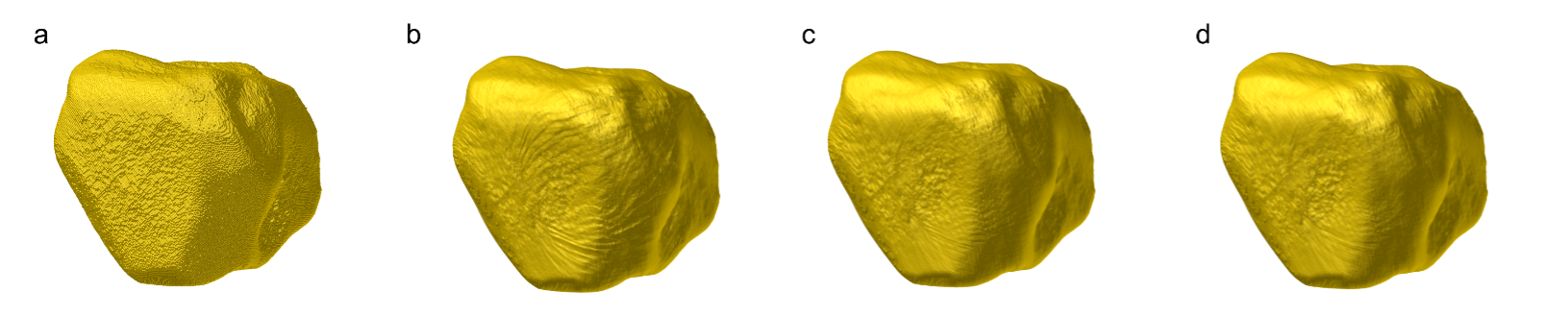}
	\caption{3D reference structure (a) along with NC-1 collected with 22 (b), 55 (c) and 71 (d) projections.}
    \label{fig:s6}
\end{figure*}

\begin{figure*}[!htb]
	\centering
	\includegraphics[width=0.5\textwidth]{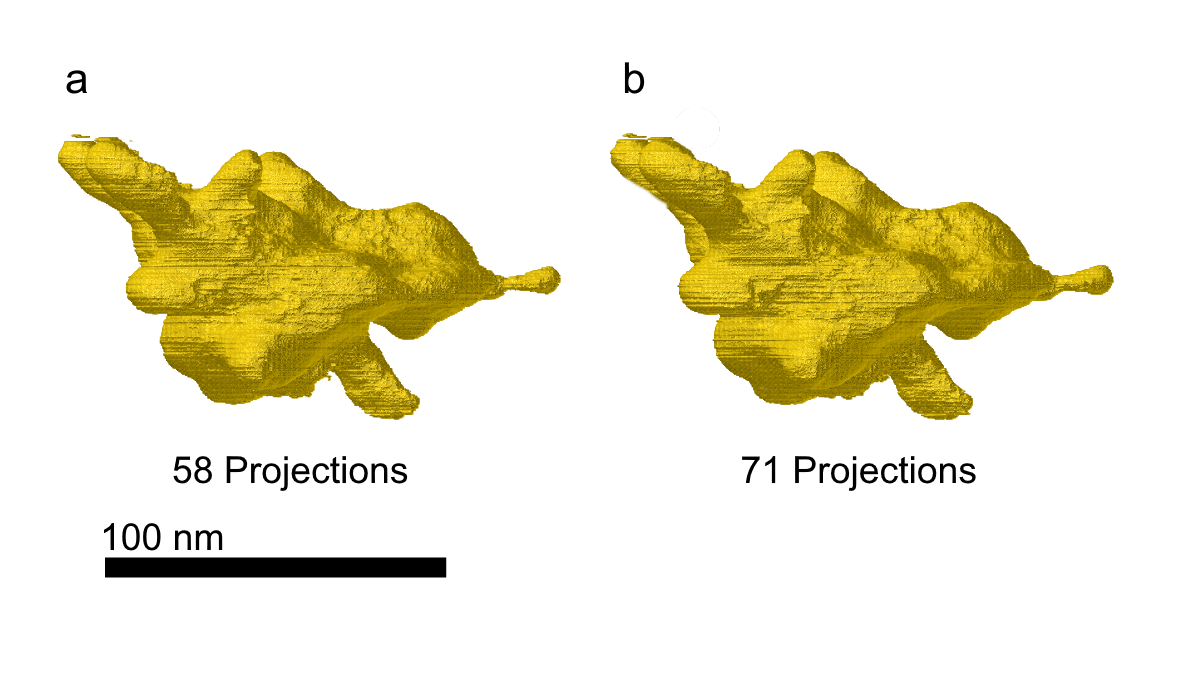}
	\caption{3D reconstruction of a Au/Pd nanostar acquired with GRS using 59 (a) and 71 projections (b).}
    \label{fig:s7}
\end{figure*}

\begin{figure*}[!htb]
	\centering
	\includegraphics[width=\textwidth]{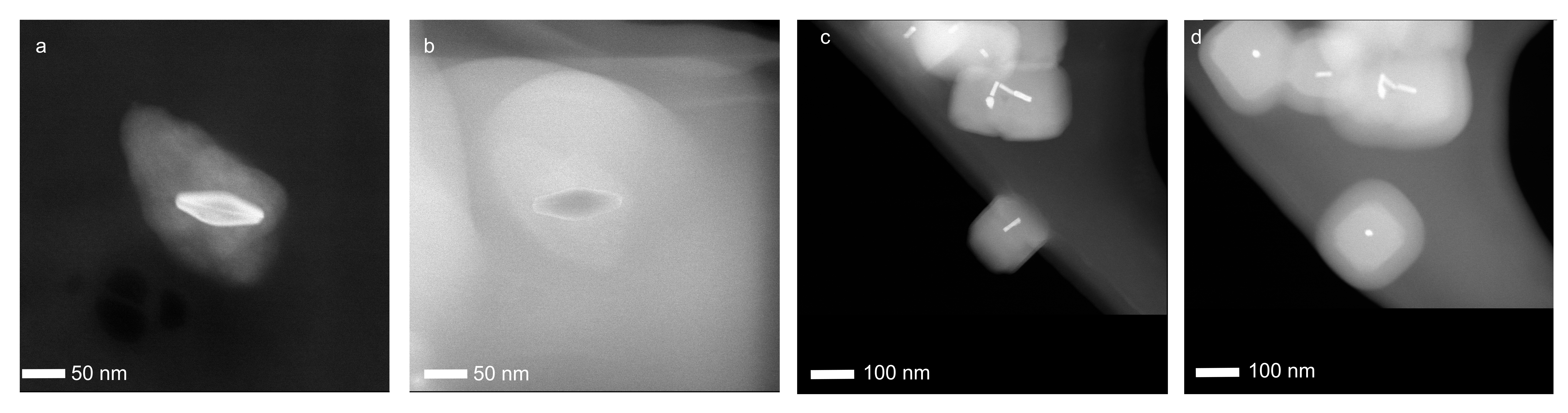}
	\caption{Electron microscopy image of Au@NU-100 collected before (a) and after (b) GRS acquisition along with before (c) and after (d) images collected for the GRS acquisition of Au/Pd@ZIF-8.}
    \label{fig:s8}
\end{figure*}

\begin{figure*}[!htb]
	\centering
	\includegraphics[width=0.6\textwidth]{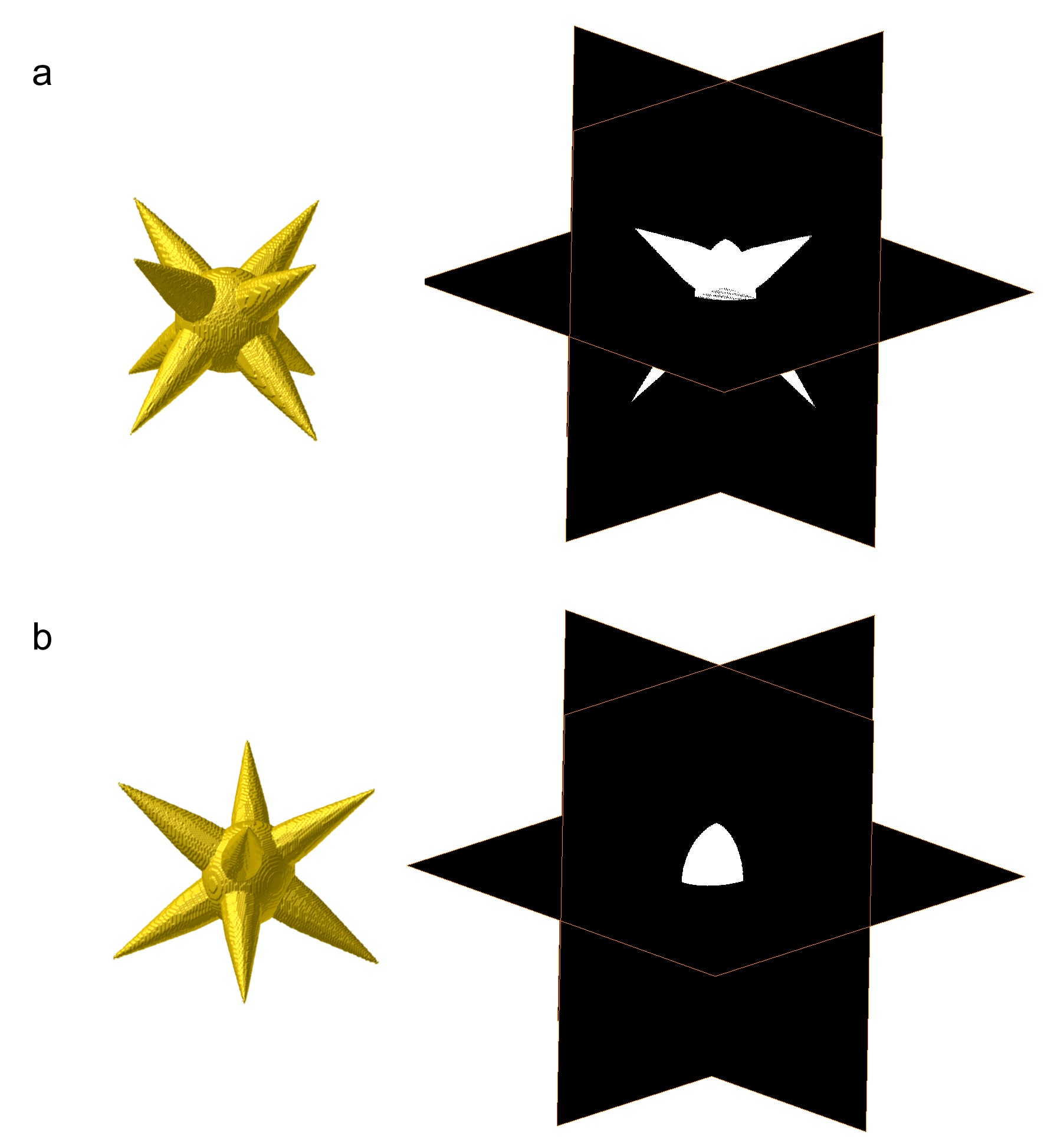}
	\caption{Nanostar volume and xy, yz, and xz orthoslices (a) prior to a 45° rotation around the y-axis (b). Dendrites that were apparent in the structure before rotation are no longer visible.}
    \label{fig:s9}
\end{figure*}


\begin{thebibliography}{10}

\bibitem{aulakh2019direct}
Darpandeep Aulakh, Lingmei Liu, Juby~R Varghese, Haomiao Xie, Timur Islamoglu,
  Kyle Duell, Chung-Wei Kung, Chia-En Hsiung, Yuxin Zhang, Riki~J Drout, et~al.
\newblock Direct imaging of isolated single-molecule magnets in metal--organic
  frameworks.
\newblock {\em Journal of the American Chemical Society}, 141(7):2997--3005,
  2019.

\bibitem{banhart2006irradiation}
F~Banhart.
\newblock Irradiation of carbon nanotubes with a focused electron beam in the
  electron microscope.
\newblock {\em Journal of materials science}, 41(14):4505--4511, 2006.

\bibitem{barreto2011nanomaterials}
Jos{\'e}~A Barreto, William O’Malley, Manja Kubeil, Bim Graham, Holger
  Stephan, and Leone Spiccia.
\newblock Nanomaterials: applications in cancer imaging and therapy.
\newblock {\em Advanced materials}, 23(12):H18--H40, 2011.

\bibitem{bartesaghi2008classification}
Alberto Bartesaghi, P~Sprechmann, J~Liu, G~Randall, G~Sapiro, and Sriram
  Subramaniam.
\newblock Classification and 3d averaging with missing wedge correction in
  biological electron tomography.
\newblock {\em Journal of structural biology}, 162(3):436--450, 2008.

\bibitem{batenburg2011dart}
Kees~Joost Batenburg and Jan Sijbers.
\newblock Dart: a practical reconstruction algorithm for discrete tomography.
\newblock {\em IEEE Transactions on Image Processing}, 20(9):2542--2553, 2011.

\bibitem{buurlage2018real}
Jan-Willem Buurlage, Holger Kohr, Willem~Jan Palenstijn, and K~Joost Batenburg.
\newblock Real-time quasi-3d tomographic reconstruction.
\newblock {\em Measurement Science and Technology}, 29(6):064005, 2018.

\bibitem{buurlage2019real}
Jan-Willem Buurlage, Federica Marone, Dani{\"e}l~M Pelt, Willem~Jan Palenstijn,
  Marco Stampanoni, K~Joost Batenburg, and Christian~M Schlep{\"u}tz.
\newblock Real-time reconstruction and visualisation towards dynamic feedback
  control during time-resolved tomography experiments at tomcat.
\newblock {\em Scientific Reports}, 9(1):1--11, 2019.

\bibitem{calvaresi2020route}
Matteo Calvaresi.
\newblock The route towards nanoparticle shape metrology.
\newblock {\em Nature nanotechnology}, 15(7):512--513, 2020.

\bibitem{chen2018biomineralized}
Ting-Ting Chen, Jin-Tao Yi, Yan-Yan Zhao, and Xia Chu.
\newblock Biomineralized metal--organic framework nanoparticles enable
  intracellular delivery and endo-lysosomal release of native active proteins.
\newblock {\em Journal of the American Chemical Society}, 140(31):9912--9920,
  2018.

\bibitem{cho2012feasibility}
Seungryong Cho, Taewon Lee, Jonghwan Min, and Hyekyun Chung.
\newblock Feasibility study on many-view under-sampling technique for low-dose
  computed tomography.
\newblock {\em Optical Engineering}, 51(8):080501, 2012.

\bibitem{choo2021nanoparticle}
Priscilla Choo, Tingting Liu, and Teri~W Odom.
\newblock Nanoparticle shape determines dynamics of targeting nanoconstructs on
  cell membranes.
\newblock {\em Journal of the American Chemical Society}, 143(12):4550--4555,
  2021.

\bibitem{crespi1996anisotropic}
Vincent~H Crespi, Nasreen~G Chopra, Marvin~L Cohen, A~Zettl, and Steven~G
  Louie.
\newblock Anisotropic electron-beam damage and the collapse of carbon
  nanotubes.
\newblock {\em Physical review B}, 54(8):5927, 1996.

\bibitem{crewe1968high}
AV~Crewe, J~Wall, and LM~Welter.
\newblock A high-resolution scanning transmission electron microscope.
\newblock {\em Journal of Applied Physics}, 39(13):5861--5868, 1968.

\bibitem{davison1983ill}
Mark~E Davison.
\newblock The ill-conditioned nature of the limited angle tomography problem.
\newblock {\em SIAM Journal on Applied Mathematics}, 43(2):428--448, 1983.

\bibitem{dudgeon1984multidimensional}
Dan~E Dudgeon and Russell~M Mersereau.
\newblock {\em Multidimensional digital signal processing}.
\newblock Prentice-Hall, 1984.

\bibitem{egerton2012mechanisms}
RF~Egerton.
\newblock Mechanisms of radiation damage in beam-sensitive specimens, for tem
  accelerating voltages between 10 and 300 kv.
\newblock {\em Microscopy research and technique}, 75(11):1550--1556, 2012.

\bibitem{frikel2013characterization}
J{\"u}rgen Frikel and Eric~Todd Quinto.
\newblock Characterization and reduction of artifacts in limited angle
  tomography.
\newblock {\em Inverse Problems}, 29(12):125007, 2013.

\bibitem{gordon1970algebraic}
Richard Gordon, Robert Bender, and Gabor~T Herman.
\newblock Algebraic reconstruction techniques (art) for three-dimensional
  electron microscopy and x-ray photography.
\newblock {\em Journal of theoretical Biology}, 29(3):471--481, 1970.

\bibitem{goris2014monitoring}
Bart Goris, Lakshminarayana Polavarapu, Sara Bals, Gustaaf Van~Tendeloo, and
  Luis~M Liz-Marz{\'a}n.
\newblock Monitoring galvanic replacement through three-dimensional
  morphological and chemical mapping.
\newblock {\em Nano letters}, 14(6):3220--3226, 2014.

\bibitem{goris2012electron}
Bart Goris, Wouter Van~den Broek, Kees~Joost Batenburg, H~Heidari Mezerji, and
  Sara Bals.
\newblock Electron tomography based on a total variation minimization
  reconstruction technique.
\newblock {\em Ultramicroscopy}, 113:120--130, 2012.

\bibitem{heymann2022progressive}
J~Bernard Heymann.
\newblock The progressive spectral signal-to-noise ratio of cryo-electron
  micrograph movies as a tool to assess quality and radiation damage.
\newblock {\em Computer Methods and Programs in Biomedicine}, 220:106799, 2022.

\bibitem{hunter2007matplotlib}
John~D Hunter.
\newblock Matplotlib: A 2d graphics environment.
\newblock {\em Computing in science \& engineering}, 9(03):90--95, 2007.

\bibitem{jiang2015electron}
Nan Jiang.
\newblock Electron beam damage in oxides: a review.
\newblock {\em Reports on Progress in Physics}, 79(1):016501, 2015.

\bibitem{kaestner2011spatiotemporal}
Anders~P Kaestner, Beat Munch, and Pavel Trtik.
\newblock Spatiotemporal computed tomography of dynamic processes.
\newblock {\em Optical Engineering}, 50(12):123201, 2011.

\bibitem{kubel2005recent}
Christian K{\"u}bel, Andreas Voigt, Remco Schoenmakers, Max Otten, David Su,
  Tan-Chen Lee, Anna Carlsson, and John Bradley.
\newblock Recent advances in electron tomography: Tem and haadf-stem tomography
  for materials science and semiconductor applications.
\newblock {\em Microscopy and Microanalysis}, 11(5):378--400, 2005.

\bibitem{lange1984reconstruction}
Kenneth Lange, Richard Carson, et~al.
\newblock Em reconstruction algorithms for emission and transmission
  tomography.
\newblock {\em J Comput Assist Tomogr}, 8(2):306--16, 1984.

\bibitem{laux2018nanomaterials}
Peter Laux, Jutta Tentschert, Christian Riebeling, Albert Braeuning, Otto
  Creutzenberg, Astrid Epp, Val{\'e}rie Fessard, Karl-Heinz Haas, Andrea Haase,
  Kerstin Hund-Rinke, et~al.
\newblock Nanomaterials: certain aspects of application, risk assessment and
  risk communication.
\newblock {\em Archives of toxicology}, 92(1):121--141, 2018.

\bibitem{liu2020bulk}
Lingmei Liu, Daliang Zhang, Yihan Zhu, and Yu~Han.
\newblock Bulk and local structures of metal--organic frameworks unravelled by
  high-resolution electron microscopy.
\newblock {\em Communications Chemistry}, 3(1):1--14, 2020.

\bibitem{mcmullan2014comparison}
G~McMullan, AR~Faruqi, D~Clare, and R~Henderson.
\newblock Comparison of optimal performance at 300 kev of three direct electron
  detectors for use in low dose electron microscopy.
\newblock {\em Ultramicroscopy}, 147:156--163, 2014.

\bibitem{midgley2009electron}
Paul~A Midgley and Rafal~E Dunin-Borkowski.
\newblock Electron tomography and holography in materials science.
\newblock {\em Nature materials}, 8(4):271--280, 2009.

\bibitem{otsu1979threshold}
Nobuyuki Otsu.
\newblock A threshold selection method from gray-level histograms.
\newblock {\em IEEE transactions on systems, man, and cybernetics},
  9(1):62--66, 1979.

\bibitem{palenstijn2011performance}
Willem~Jan Palenstijn, K~Joost Batenburg, and Jan Sijbers.
\newblock Performance improvements for iterative electron tomography
  reconstruction using graphics processing units (gpus).
\newblock {\em Journal of structural biology}, 176(2):250--253, 2011.

\bibitem{pryor2017genfire}
Alan Pryor, Yongsoo Yang, Arjun Rana, Marcus Gallagher-Jones, Jihan Zhou,
  Yuan~Hung Lo, Georgian Melinte, Wah Chiu, Jose~A Rodriguez, and Jianwei Miao.
\newblock Genfire: A generalized fourier iterative reconstruction algorithm for
  high-resolution 3d imaging.
\newblock {\em Scientific reports}, 7(1):1--12, 2017.

\bibitem{roduner2006size}
Emil Roduner.
\newblock Size matters: why nanomaterials are different.
\newblock {\em Chemical Society Reviews}, 35(7):583--592, 2006.

\bibitem{rosler2014encapsulation}
Christoph R{\"o}sler, Daniel Esken, Christian Wiktor, Hirokazu Kobayashi,
  Tomokazu Yamamoto, Syo Matsumura, Hiroshi Kitagawa, and Roland~A Fischer.
\newblock Encapsulation of bimetallic nanoparticles into a metal--organic
  framework: Preparation and microstructure characterization of pd/au@ zif-8.
\newblock {\em European Journal of Inorganic Chemistry}, 2014(32):5514--5521,
  2014.

\bibitem{russo2018charge}
Christopher~J Russo and Richard Henderson.
\newblock Charge accumulation in electron cryomicroscopy.
\newblock {\em Ultramicroscopy}, 187:43--49, 2018.

\bibitem{scott2012electron}
MC~Scott, Chien-Chun Chen, Matthew Mecklenburg, Chun Zhu, Rui Xu, Peter Ercius,
  Ulrich Dahmen, BC~Regan, and Jianwei Miao.
\newblock Electron tomography at 2.4-{\aa}ngstr{\"o}m resolution.
\newblock {\em Nature}, 483(7390):444--447, 2012.

\bibitem{shin1989annular}
DH~Shin, EJ~Kirkland, and J~Silcox.
\newblock Annular dark field electron microscope images with better than 2
  {\aa} resolution at 100 kv.
\newblock {\em Applied physics letters}, 55(23):2456--2458, 1989.

\bibitem{treacy1987electron}
MMJ Treacy and JM~Newsam.
\newblock Electron beam sensitivity of zeolite l.
\newblock {\em Ultramicroscopy}, 23(3-4):411--419, 1987.

\bibitem{turner2008direct}
Stuart Turner, Oleg~I Lebedev, Felicitas Schroder, Daniel Esken, Roland~A
  Fischer, and Gustaaf~Van Tendeloo.
\newblock Direct imaging of loaded metal- organic framework materials (metal@
  mof-5).
\newblock {\em Chemistry of Materials}, 20(17):5622--5627, 2008.

\bibitem{ugurlu2011radiolysis}
O~Ugurlu, J~Haus, AA~Gunawan, MG~Thomas, S~Maheshwari, M~Tsapatsis, and
  KA~Mkhoyan.
\newblock Radiolysis to knock-on damage transition in zeolites under electron
  beam irradiation.
\newblock {\em Physical Review B}, 83(11):113408, 2011.

\bibitem{urban2008studying}
Knut~W Urban.
\newblock Studying atomic structures by aberration-corrected transmission
  electron microscopy.
\newblock {\em Science}, 321(5888):506--510, 2008.

\bibitem{van2016fast}
Wim Van~Aarle, Willem~Jan Palenstijn, Jeroen Cant, Eline Janssens, Folkert
  Bleichrodt, Andrei Dabravolski, Jan De~Beenhouwer, K~Joost Batenburg, and Jan
  Sijbers.
\newblock Fast and flexible x-ray tomography using the astra toolbox.
\newblock {\em Optics express}, 24(22):25129--25147, 2016.

\bibitem{van2015astra}
Wim Van~Aarle, Willem~Jan Palenstijn, Jan De~Beenhouwer, Thomas Altantzis, Sara
  Bals, K~Joost Batenburg, and Jan Sijbers.
\newblock The astra toolbox: A platform for advanced algorithm development in
  electron tomography.
\newblock {\em Ultramicroscopy}, 157:35--47, 2015.

\bibitem{vanrossum}
Guido Van~Rossum and Fred~L. Drake.
\newblock {\em Python 3 Reference Manual}.
\newblock CreateSpace, Scotts Valley, CA, 2009.

\bibitem{vanrompay2019experimental}
Hans Vanrompay, Armand B{\'e}ch{\'e}, Johan Verbeeck, and Sara Bals.
\newblock Experimental evaluation of undersampling schemes for electron
  tomography of nanoparticles.
\newblock {\em Particle \& Particle Systems Characterization}, 36(7):1900096,
  2019.

\bibitem{vanrompay20183d}
Hans Vanrompay, Eva Bladt, Wiebke Albrecht, Armand B{\'e}ch{\'e}, Marina
  Zakhozheva, Ana S{\'a}nchez-Iglesias, Luis~M Liz-Marz{\'a}n, and Sara Bals.
\newblock 3d characterization of heat-induced morphological changes of au
  nanostars by fast in situ electron tomography.
\newblock {\em Nanoscale}, 10(48):22792--22801, 2018.

\bibitem{vanrompay2020real}
Hans Vanrompay, Jan-Willem Buurlage, Dani{\"e}l~M Pelt, Vished Kumar, Xiaolu
  Zhuo, Luis~M Liz-Marz{\'a}n, Sara Bals, and K~Joost Batenburg.
\newblock Real-time reconstruction of arbitrary slices for quantitative and in
  situ 3d characterization of nanoparticles.
\newblock {\em Particle \& Particle Systems Characterization}, 37(7):2000073,
  2020.

\bibitem{vanrompay2021fast}
Hans Vanrompay, Alexander Skorikov, Eva Bladt, Armand B{\'e}ch{\'e}, Bert
  Freitag, Johan Verbeeck, and Sara Bals.
\newblock Fast versus conventional haadf-stem tomography of nanoparticles:
  advantages and challenges.
\newblock {\em Ultramicroscopy}, 221:113191, 2021.

\bibitem{zak2021guide}
Andrzej {\.Z}ak.
\newblock Guide to controlling the electron dose to improve low-dose imaging of
  sensitive samples.
\newblock {\em Micron}, 145:103058, 2021.

\end{thebibliography}
\end{document}